\documentclass[fleqn,10pt]{wlscirep}
\usepackage[utf8]{inputenc}
\usepackage[T1]{fontenc}

\usepackage{booktabs}
\usepackage{multirow}
\usepackage{pdflscape}
\usepackage{array}
\usepackage{ragged2e}
\usepackage{afterpage}
\usepackage{tabularx}
\usepackage{makecell}
\usepackage{longtable}
\usepackage[normalem]{ulem}
\useunder{\uline}{\ul}{}
\usepackage[most]{tcolorbox}

\renewcommand{\arraystretch}{1.2}
\newcolumntype{P}[1]{>{\raggedright\arraybackslash}p{#1}}
\newcolumntype{C}[1]{>{\centering\arraybackslash}p{#1}}
\usepackage{pdflscape}
\usepackage{afterpage}

\newcommand{\rom}[1]{\uppercase\expandafter{\romannumeral #1\relax}}

\title{AI summaries in online search influence users' attitudes}

\author[1*]{Yiwei Xu}
\author[2]{Saloni Dash}
\author[1]{Sungha Kang}
\author[3]{Wang Liao}
\author[2]{Emma S. Spiro}
\affil[1]{University of Maryland, College of Information, College Park, MD 20742, United States}
\affil[2]{University of Washington, Information School, Seattle, WA 98195, United States.}
\affil[3]{University of Washington, Department of Communication, Seattle, WA 98195, United States.}

\affil[*]{yiweixu@umd.edu}

\begin{abstract}
This study examined how AI-generated summaries, which have become visually prominent in online search results, affect how users think about publicly debated issues. In a preregistered randomized controlled experiment, participants (\textit{N} = 2,004) viewed mock search result pages varying in the presence (vs. absence), placement (top vs. middle), and stance (benefit-framed vs. harm-framed) of AI-generated summaries across four publicly debated topics. Compared to a no-summary control group, participants exposed to AI-generated summaries reported issue attitudes, behavioral intentions, and policy support that aligned more closely with the AI summary stance. The summaries placed at the top of the page produced stronger shifts in users' issue attitudes (but not behavioral intentions or policy support) than those placed at the middle of the page. We also observed moderating effects from issue familiarity and general trust toward AI. In addition, users perceived the AI summaries more useful when it emphasized health harms versus benefits. These findings suggest that AI-generated search summaries can significantly shape public perceptions, raising important implications for the design and regulation of AI-integrated information ecosystems.
\par\vspace{1em}


\par\vspace{1em}

\textbf{Significance statement:}
This study provides some of the first causal evidence on how AI-generated summaries in online search results can meaningfully influence users’ issue attitudes, as well as behavioral intentions and policy support for publicly debated issues. This work establishes important baseline evidence for future studies of AI-driven information seeking and persuasion in more interactive and ecologically valid settings. The findings also inform AI developers, healthcare providers, and policymakers in designing responsible AI-integrated information systems that support equitable and trustworthy online information seeking and processing.

\end{abstract}

\begin{document}
\flushbottom
\maketitle
%
%
\thispagestyle{empty}


\section*{Introduction}

People increasingly rely on online search for timely, low-cost information and advice across a wide range of topics. Among the tools that support online information seeking, generative artificial intelligence (AI) has quickly become prominent. As major search engines such as Google and Bing integrate AI-driven features into their interfaces, a growing share of the public now encounters AI-generated summaries, recommendations, and synthesized responses as part of everyday search \cite{eddy_americans_2025}. These integrations have the potential to fundamentally reshape how users process, navigate, and evaluate online information \cite{memon_search_2024}. For example, AI-generated search summaries can present tailored, fluent answers synthesized from (seemingly) diverse sources, helping users quickly extract key insights or arrive at targeted conclusions. As AI-mediated information seeking becomes prevalent, researchers must examine how these tools influence users’ evaluations, judgments, and decision-making about complex or contested issues. This is the question we investigate in this study. 
  
A growing body of research shows that AI-generated content can meaningfully shape user beliefs and attitudes. For example, AI-generated or AI-paraphrased political messages can be as persuasive as, or even more persuasive than, human-written messages \cite{goldstein_how_2024, dash_persuasive_2025}. When communicating about public health issues, AI-generated messages have also outperformed original institutional communications in persuasiveness \cite{karinshak_working_2023}. AI can also personalize messages which are significantly more persuasive than non-personalized ones across domains such as marketing and politics \cite{matz_potential_2024}. When AI is used as writing assistants, biased autocomplete suggestions can shift users’ attitudes toward the AI’s position \cite{williams-ceci_biased_2025}. Researchers also found that opinionated LLM can reinforce users' views and exacerbate their bias \cite{sharma_generative_2024}. AI-generated smart replies can alter interpersonal perceptions and dynamics \cite{mieczkowski_ai-mediated_2021}. Together, this body of work reveals that AI-generated content, whether fully composed or interactively integrated, can influence user beliefs. Similarly, AI-generated outputs may demonstrate persuasive effects in other everyday information environments, such as search engines integrated with AI tools. Although the specific ways users rely on these AI-generated summaries is not yet fully understood, it is possible that these summaries, which can simplify complex search results and are often presented first on a search results page, could serve as an information anchor and influence how users form judgments and make decisions on the topics they search. To the best of our knowledge, we are the first to experimentally evaluate the effects of AI summaries in online search on users’ attitudes. Here, we discuss the theoretical grounding and motivation for the hypotheses that structure this study. 

\subsection*{The Presence Versus Absence of AI Summary in Online Search}
Some theories of information processing, such as the heuristic–systematic model (HSM) \cite{eagly_psychology_1993}, suggest that users often rely on cognitive shortcuts or accessible cues when forming judgments under limited attention to minimize the use of cognitive resources and to conserve mental effort. As such, when seeking and processing information online through AI-integrated search engines, users are likely to rely on similar shortcuts to quickly process search results and form judgments. 

Summarized texts, when compared to full texts, have been found to help people recall information and learn knowledge better, and to improve knowledge transfer to related domains more effectively \cite{reder_comparison_1980}. Summarized information may not only aid learning but also influence attitudes and beliefs. This is because AI-generated summaries may function as heuristic cues by simplifying complex information or debates into accessible takeaways and facilitating the evaluation of information with less effort. As a result, AI-generated online search summaries may shift user beliefs toward the stance they advocate, especially in the absence of deep elaboration. We hypothesize the following:

Preregistered Hypothesis 1a, 1b, 1c: The presence (vs. absence) of an AI-generated summary influences user beliefs by moving their (a) issue attitudes, (b) behavioral intentions, (c) policy attitudes closer to the stance advocated in the summary.

In this study, we focus on how AI summaries affect three belief-based outcomes central to persuasion: (1) issue attitudes, or users’ overall evaluations of the topic; (2) behavioral intentions, or their stated likelihood of performing related behaviors; and (3) policy attitudes, or their support for relevant public policies. These outcomes reflect important cognitive and behavioral targets often shaped by persuasive messaging and are particularly relevant in decision-making.

\subsection*{Anchoring and Primacy Effects}

Heuristics (defined as a collection of cognitive processes, either conscious or unconscious) impact assessment, recall, and use of information in decision making. A widely used and powerful heuristic, known as anchoring, describes a user’s tendency to over rely on the first piece of information they receive when making a decision. Anchoring significantly impacts the user’s subsequent choices \cite{tversky_judgment_1974}. For example, researchers found that a sponsored search result, which is ranked as the first item on the search result webpage, influences subsequent consumer choice \cite{bhattacharya_examining_2024}. A related heuristic is primacy. Primacy effects occur when users are more likely to remember the information presented first and base their decisions on this primary information \cite{brunel_message_2003}. Both anchoring and primacy effects suggest an outsized impact of the initial information on people's decision making. AI-generated search engine summaries, consistently positioned as the first piece of information on a search results page, may evoke these heuristics, function as an anchor, and prime users' judgments of the issue they search. Based on this, we hypothesize the following:

Preregistered Hypothesis 2a, 2b, 2c: When the AI-generated summary is present, its position on the page (top vs. middle) influences user beliefs through anchoring, with top positioning exerting a stronger persuasive effect by moving their (a) issue attitudes, (b) behavioral intentions, (c) policy attitudes closer to the stance advocated in the summary.

\subsection*{Message Level and User Level Moderators}

In addition to the two main hypotheses (H1 and H2), we also examine potential moderators -- issue stance of the AI summary, users' issue familiarity, and general trust in AI -- to assess whether these factors moderate the persuasive effects of AI-generated summaries. 

We first explore whether AI-generated summaries emphasizing health \textit{harms} exert different levels of persuasive influence than those emphasizing \textit{benefits} (i.e., issue stance). Gain–loss framing theory suggests that the effectiveness of a message varies; loss-framed messages (emphasizing harms) often eliciting stronger responses when behaviors involve perceived risk or uncertainty \cite{rothman_shaping_1997}. In contrast, gain-framed messages (emphasizing benefits) prove more effective for promoting certain health behaviors such as prevention \cite{gallagher_health_2012, okeefe_relative_2007}. We explore whether the AI summary's issue stance moderates the effects of its presence and placement on the persuasive outcomes (Research Question 1).

Next, we explore user-level moderators that may shape the persuasive impact of AI-generated summaries.
Individuals’ familiarity with an issue (i.e., issue familiarity) may influence how they respond to online content. Persuasive messages are generally more effective when the content is perceived as novel, since unfamiliar arguments are more likely to prompt opinion change \cite{sears_effects_1965}. This suggests that users with lower issue familiarity may be more influenced by persuasive content, prompting us to examine whether issue familiarity moderates the effects of AI summaries (Research Question 2a).
We also considered users’ general trust in AI as another potential moderator, given that trust in the information source can shape message acceptance \cite{hohenberg_when_2023}. We examined whether general trust in AI moderates the influence of AI-generated summaries (Research Question 2b).

Together, these moderators help identify the boundary conditions of AI-generated summaries’ persuasive influence in online information search. 

\subsection*{User Experience Related Outcomes}
In addition, we are interested in how users perceive the AI-generated summaries themselves, including AI summaries' trustworthiness and usefulness, and their intention in keeping or disabling such function in the future. These perceptions can inform future research on human-AI engagement, as well as tool design. Specifically, we explored how the issue direction and position of an AI-generated summary influence users’ evaluations of the summary itself (Research Question 3), offering a more holistic view of user–AI interactions in the context of online search.

\subsection*{The Present Research}
This study aims to provide some of the first empirical evidence regarding the social and psychological impact of AI-integrated information-seeking tools. We conducted a preregistered randomized controlled online experiment (\textit{N} = 2,004) to explore whether the presence (Preregistered H1) and the placement (Preregistered H2) of AI-generated summaries that appear in search results affect users' (a) issue attitudes, (b) behavioral intentions, and (c) policy support about different publicly debated health issues. We also explored the potential moderating effects of issue stance (RQ1), user issue familiarity (RQ2a), and user trust (RQ2b), along with the effects of presence and placement on user experience (RQ3). The full hypotheses and research questions are detailed in the Methods section. The study and its two primary hypotheses were preregistered at Open Science Framework (OSF): \url{https://osf.io/n9e7x/?view_only=834769ba56094c3cbbc236f353e301c0}

This work makes several important contributions: It provides some of the first robust empirical evidence on the causal effects of AI-generated search summaries, showing that AI summaries on the search results page can meaningfully influence users’ issue attitudes, behavioral intentions and policy support. This work also extends persuasion and information processing theories such as anchoring and primacy effects to examine the effects of AI-integrated online information seeking tools. This work establishes important baseline evidence for future studies of AI-assisted information seeking and persuasion in more interactive and ecologically valid settings across a range of issue domains.

Research of this kind aims to enable AI developers, public health organizations, and policymakers to improve the safe application of generative AI for people’s information seeking, processing, and decision making. Importantly, populations with limited access to trusted professional expertise may be especially reliant on these emerging tools, making it increasingly important to understand and safeguard AI tools' influence on public decision-making.

\section*{Results}



\subsection*{The Presence (vs. Absence) of an AI-generated Summary Influenced Users' Issue Attitudes, Behavioral Intentions, and Policy Support (H1a, H1b, H1c)}

Our analysis (see Table \ref{hypothesis} and Figure \ref{fig:outcome_condition}) revealed that the presence of an AI-generated summary on a search results page, compared to a control condition without any AI summary, significantly influenced users’ issue attitudes (\textit{F}(2, 1995) = 54.57, \textit{p} $<$ .001, $\eta_{p}^{2}$ = .052; H1a supported), behavioral intentions (\textit{F}(2, 1995) = 18.55, \textit{p} $<$ .001, $\eta_{p}^{2}$ = .018; H1b supported), and policy support (\textit{F}(2, 1995) = 16.76, \textit{p} $<$ .001, $\eta_{p}^{2}$ = .017; H1c supported), making these outcomes more aligned with the AI summary’s stance.

\subsection*{The Placement (Top vs. Middle) of an AI-generated Summary Influenced Users' Issue Attitudes (H2a, H2b, H2c)}

\begin{figure} 
	\centering
	\includegraphics[width=1\textwidth]{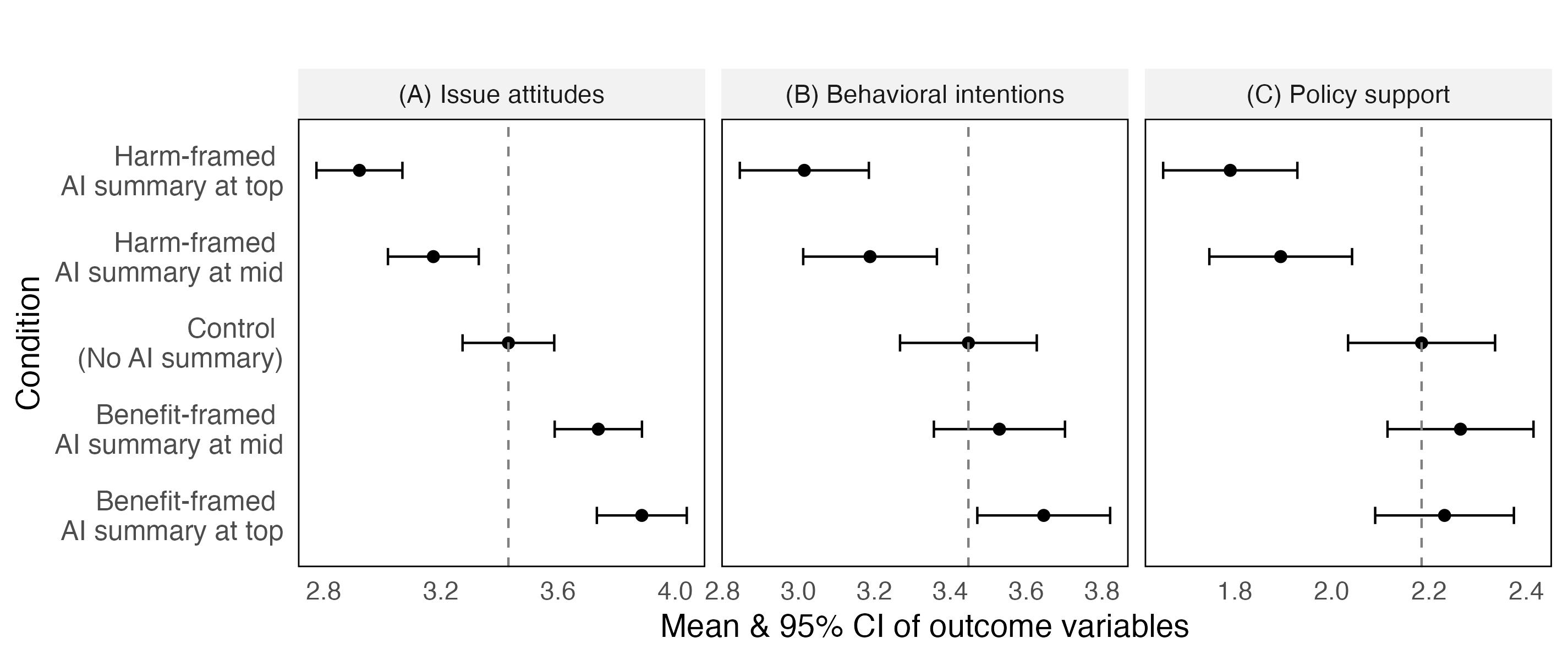}

	\caption{\textbf{Main outcome variables by condition.}
        (\textbf{A}) Issue attitudes by condition. (\textbf{B}) Behavioral intentions by condition. (\textbf{C}) Policy support by condition.
        }
	\label{fig:outcome_condition} 
\end{figure}

Our analysis (see Table \ref{hypothesis} and Figure \ref{fig:outcome_condition}) revealed that a top-placed (vs. mid-placed) AI summary significantly influenced users' issue attitudes, \textit{F}(2, 1993) = 3.87, \textit{p} = .021, $\eta_{p}^{2}$ = .004 (H2a supported). However, no significant effects of AI summary placement (top vs. mid) were found on behavioral intentions (\textit{F}(2, 1993) = 1.64, \textit{p} = .195, $\eta_{p}^{2}$ = .002; H2b not supported) or policy support (\textit{F}(2, 1993) = 0.536, \textit{p} = .585, $\eta_{p}^{2}$ = .001; H2c not supported). 

\begin{table}[ht!]
\small
\centering
\renewcommand{\arraystretch}{1.0}
\setlength{\tabcolsep}{4pt}
\caption{\textbf{Hypothesis Testing (with Robust SE)}}
\label{hypothesis}
\begin{tabular}{C{3.8cm} C{3.3cm} | C{1cm} C{1cm} C{1cm} C{1.1cm} C{1cm} | C{3.6cm}}
\toprule
\textbf{Outcomes / Conditions}
& \textbf{Joint Tests of Planned Contrasts} 
& \multicolumn{5}{>{\centering\arraybackslash}p{5.5cm}|}{\textbf{Planned Contrasts (Holm--Bonferroni Adjusted)}} 
& \textbf{Contrasts of Planned Contrasts} \\
& \textit{F (df1, df2), p, d} 
& \textit{Est.} & \textit{SE} & \textit{t} & \textit{p} & \textit{d} 
& \textit{t (df), p, d} \\
\midrule

\textbf{Issue Attitudes} & \textbf{PH1a (supported)} & & & & & & \textbf{RQ1} \\
Benefit-framed AI Summaries vs. Control & \multirow{2}{=}{$F(2, 1995) = 54.57, p < .001, \eta_p^2 = .052$} 
& 0.375 & 0.091 & 4.117 & $<.001$ & 0.253 & \multirow{2}{=}{$t(1995) = -0.14, p = .891, d = 0.016$} \\
Harm-framed AI Summaries vs. Control & & -0.398 & 0.092 & -4.324 & $<.001$ & -0.268 & \\

\textbf{Behavioral Intentions} & \textbf{PH1b (supported)} & & & & & & \textbf{RQ1} \\
Benefit-framed AI Summaries vs. Control & \multirow{2}{=}{$F(2, 1995) = 18.55, p < .001, \eta_p^2 = .018$} 
& 0.146 & 0.100 & 1.456 & 0.146 & 0.089 & \multirow{2}{=}{$t(1995) = -1.10, p = .273, d = -0.123$} \\
Harm-framed AI Summaries vs. Control & & -0.347 & 0.101 & -3.444 & 0.001 & -0.212 & \\

\textbf{Policy Support} & \textbf{PH1c (supported)} & & & & & & \textbf{RQ1} \\
Benefit-framed AI Summaries vs. Control & \multirow{2}{=}{$F(2, 1995) = 16.76, p < .001, \eta_p^2 = .017$} 
& 0.063 & 0.092 & 0.686 & 0.493 & 0.043 & \multirow{2}{=}{$t(1995) = -1.66, p = .097, d = -0.189$} \\
Harm-framed AI Summaries vs. Control & & -0.340 & 0.091 & -3.758 & $<.001$ & -0.232 & \\

\textbf{Issue Attitudes} & \textbf{PH2a (supported)} & & & & & & \textbf{RQ1} \\
Benefit-framed AI Summary at Top vs. Middle & \multirow{2}{=}{$F(2, 1993) = 3.87, p = .021, \eta_p^2 = .004$} 
& 0.145 & 0.103 & 1.402 & 0.161 & 0.098 & \multirow{2}{=}{$t(1993) = -0.74, p = .457, d = -0.074$} \\
Harm-framed AI Summary at Top vs. Middle & & -0.255 & 0.106 & -2.401 & 0.033 & -0.172 & \\

\textbf{Behavioral Intentions} & \textbf{PH2b (not supported)} & & & & & & \textbf{RQ1} \\
Benefit-framed AI Summary at Top vs. Middle & \multirow{2}{=}{$F(2, 1993) = 1.64, p = .195, \eta_p^2 = .002$} 
& 0.123 & 0.115 & 1.064 & 0.288 & 0.075 & \multirow{2}{=}{$t(1993) = -0.30, p = .766, d = -0.03$} \\
Harm-framed AI Summary at Top vs. Middle & & -0.172 & 0.117 & -1.463 & 0.287 & -0.105 & \\

\textbf{Policy Support} & \textbf{PH2c (not supported)} & & & & & & \textbf{RQ1} \\
Benefit-framed AI Summary at Top vs. Middle & \multirow{2}{=}{$F(2, 1993) = 0.54, p = .585, \eta_p^2 = .001$} 
& -0.025 & 0.105 & -0.233 & 0.816 & -0.017 & \multirow{2}{=}{$t(1993) = -0.87, p = .386, d = -0.086$} \\
Harm-framed AI Summary at Top vs. Middle & & -0.102 & 0.101 & -1.009 & 0.626 & -0.070 & \\

\bottomrule
\end{tabular}

\caption*{\footnotesize 
\textit{Notes.} 
1. The test of H1 was conducted using a joint test (left column) of two planned contrasts (middle column) – benefit-framed AI vs. control, harm-framed AI vs. control. In addition, we conducted pairwise comparison (right column) between these two planned contrasts to test whether the effect of the presence (versus absence) of AI summary was moderated by issue stance (RQ1). 
2. The test of H2 was conducted using a joint test (left column) of two planned contrasts (middle column) – benefit-framed AI summary at top vs. benefit-framed AI summary at mid, harm-framed AI summary at top vs. harm-framed AI summary at middle. In addition, we conducted pairwise comparison between these two planned contrasts (right column) to test whether the effect of the placement (top vs. mid) of AI summary was moderated by issue stance (RQ1).
}
\end{table}


\subsection*{The Moderating Roles of Issue Stance, Issue Familiarity, and General Trust in AI}

\subsubsection*{AI Summary Issue Stance (RQ1)}

\paragraph{AI Summary Presence X Issue Stance.}
The issue stance of the AI summary (benefit-framed vs. harm-framed) did not moderate the persuasive effects of AI summary presence (vs. control) for any of the three primary outcome variables: issue attitudes (\textit{t}(1995) = -0.14, \textit{p} = .891, \textit{d} = 0.02), behavioral intentions (\textit{t}(1995) = -1.10, \textit{p} = .273, \textit{d} = -0.12), and policy support (\textit{t}(1995) = -1.66, \textit{p} = .097, \textit{d} = -0.19). The findings suggest that whether the AI summary focused on benefits or harms did not change how persuasive it was when comparing to having no summary.

\paragraph{AI Summary Placement X Issue Stance.}
The issue stance of the AI summary (benefit-framed vs. harm-framed) did not moderate the effects of AI summary placement (top vs. middle) across the three primary outcome variables: issue attitudes (\textit{t}(1993) = -0.74, \textit{p} = .457, \textit{d} = -0.07), behavioral intentions (\textit{t}(1993) = -0.30, \textit{p} = .766, \textit{d} = -0.03), and policy support (\textit{t}(1993) = -0.87, \textit{p} = .386, \textit{d} = -0.09). The findings suggest that the issue stance of the AI summary (benefit vs. harm) did not change how persuasive the summary was based on its placement (top vs. mid).

\subsubsection*{User Issue Familiarity (RQ2a)}

\begin{figure}[ht!] 
	\centering
	\includegraphics[width=0.88\textwidth]{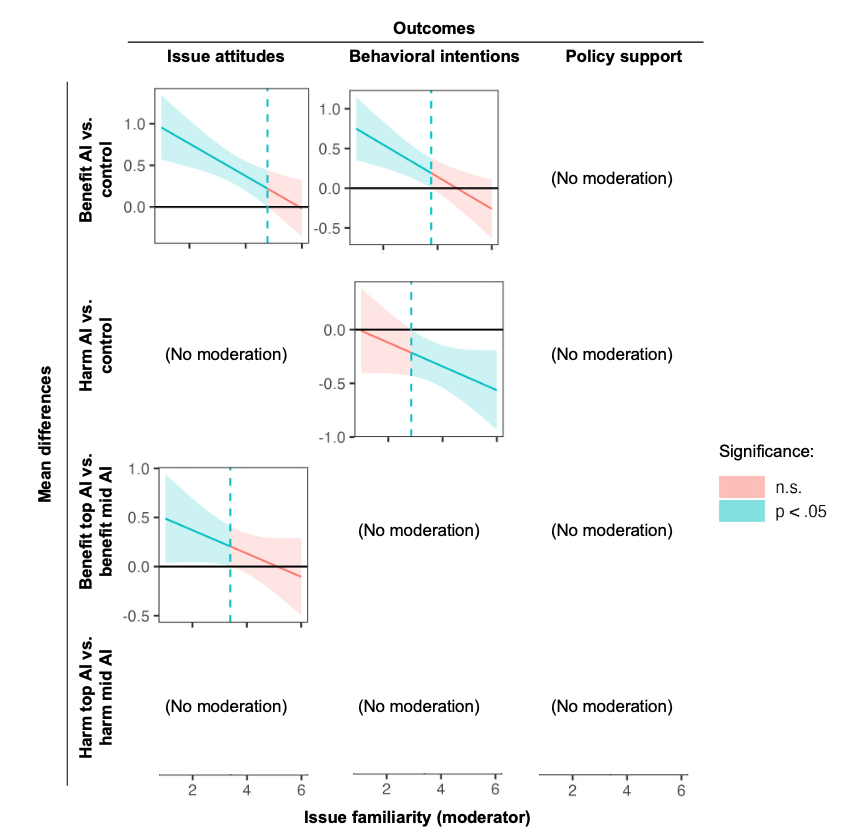} 

	\caption{\textbf{Johnson-Neyman plots with user issue familiarity as moderator.}
		Each panel shows the estimated average outcome differences between two conditions, given a certain level of issue familiarity. Non-significantly moderated comparisons are omitted.
        }
	\label{fig:JN_familiarity} 
\end{figure}

\paragraph{AI Summary Presence X Familiarity.}
The interaction effects between AI summary presence (benefit-framed AI summaries vs. control) and familiarity were (marginally) significant for issue attitudes (\textit{B} = -0.19, \textit{p} = .003), behavioral intentions (\textit{B} = -0.20, \textit{p} = .003), marginally significant for policy support (\textit{B} = -0.09, \textit{p} = .146). 
Specifically, when participants were less familiar with the issue, the presence of benefit-framed AI summaries (vs. control without AI summaries) were more persuasive (\textit{p} $<$ .05) in influencing people's issue attitudes (when familiarity $<$ 4.77) and behavioral intentions (when familiarity $<$ 3.77) (see the first row of Figure \ref{fig:JN_familiarity}).

The interaction effects between AI summary presence (harm-framed AI summaries vs. control) and familiarity were not significant for issue attitudes (\textit{p} = .630), marginally significant for behavioral intentions (\textit{B} = -0.11, \textit{p} = .103), not significant for policy support (\textit{p} = .459).
Specifically, when participants were more familiar with the issue, the presence of benefit-framed AI summaries (vs. control without AI summaries) were more persuasive (\textit{p} $<$ .05) in influencing people's behavioral intentions (when familiarity $>$ 2.85) (see the second row of Figure \ref{fig:JN_familiarity}).

\paragraph{AI Summary Placement X Familiarity.}
The interaction effects between AI summary placement (benefit-top-AI vs. benefit-mid-AI) and familiarity were (marginally) significant for issue attitudes (\textit{B} = -0.12, \textit{p} = .120), not significant for behavioral intentions (\textit{p} = .367) or for policy support (\textit{p} = .604). 
Specifically, when users were less familiar with the issue, the benefit-top (vs. benefit-mid) condition was more persuasive (\textit{p} $<$ .05) in influencing people's issue attitudes (when familiarity $<$ 3.39) (see the third row of Figure \ref{fig:JN_familiarity}). 

The interaction items between AI summary placement (harm-top-AI vs. harm-mid-AI) and familiarity were not significant for issue attitudes (\textit{p} = .615), behavioral intentions (\textit{p} = .428), policy support (\textit{p} = .499).


\subsubsection*{User Trust toward AI in General (RQ2b)}

\begin{figure}[ht!] 
	\centering
	\includegraphics[width=0.88\textwidth]{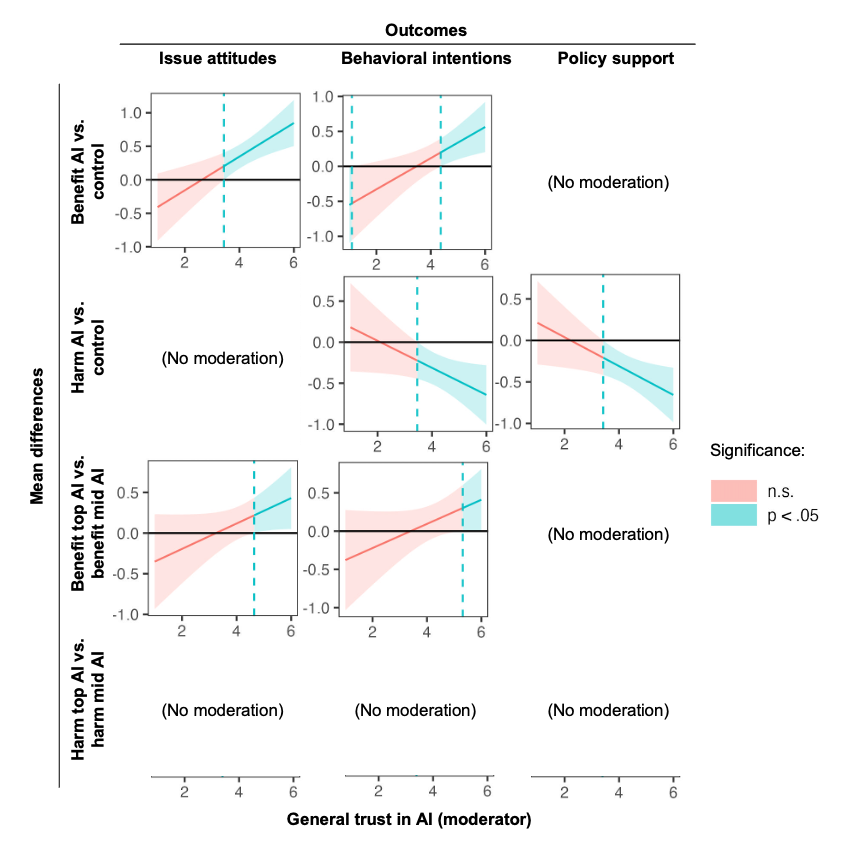} 

	\caption{\textbf{Johnson-Neyman plots with user general trust in AI as moderator}
    Each panel shows the estimated average outcome differences between two conditions, given a certain level of general trust in AI. Non-significantly moderated comparisons are omitted.
        }
	\label{fig:JN_trust} 
\end{figure}

\paragraph{AI Summary Presence X Trust.}
The interaction effects between AI summary presence (benefit-framed AI summaries vs. control) and trust were significant for issue attitudes (\textit{B} = 0.25, \textit{p} = .001), behavioral intentions (\textit{B} = 0.22, \textit{p} = .007), marginally significant for policy support (\textit{B} = 0.12, \textit{p} = .098).
Specifically, benefit-framed summaries (vs. control) were more persuasive (\textit{p} $<$ .05; see Figure) in influencing people’s issue attitudes (when trust $>$ 3.43) and behavioral intentions (when trust $>$ 4.37) (see the first row of Figure \ref{fig:JN_trust}).

The interaction items between AI summary presence (harm-framed AI summaries vs. control) and trust were not significant for issue attitudes (\textit{p} = .314), significant for behavioral intentions (\textit{B} = -0.16, \textit{p} = .045) and for policy support (\textit{B} = -0.17, \textit{p} = .023).
Specifically, when participants had more trust toward AI, harm-framed summaries (vs. control) were more persuasive (\textit{p} $<$ .05) in influencing people’s behavioral intentions (when trust $>$ 3.45) and policy support (when trust $>$ 3.42) (see the second row of Figure \ref{fig:JN_trust}).

\paragraph{AI Summary Placement X Trust.}
The interaction effects between AI summary placement (benefit-top-AI vs. benefit-mid-AI) and trust were (marginally) significant for issue attitudes (\textit{B} = 0.16, \textit{p} = .078), behavioral intentions (\textit{B} = 0.158, \textit{p} = .103), not significant for policy support (\textit{p} = .218). 
Specifically, when users had more trust in AI in general, the benefit-top (vs. benefit-mid) condition was more persuasive (\textit{p} $<$ .05) in influencing people's issue attitudes (when trust $>$ 4.64) and behavioral intentions (when trust $>$ 5.45) (see the third row of Figure \ref{fig:JN_trust}). 

The interaction effects between AI summary placement (harm-top-AI vs. harm-mid-AI) and trust were not significant for issue attitudes (\textit{p} = .924), behavioral intentions (\textit{p} = .965), and policy support (\textit{p} = .183).

\subsection*{Users Found Harm-framed (vs. Benefit-framed) AI-generated Summaries More Useful (RQ3)}


The results revealed a significant main effect of the AI summary’s issue stance on user perceived usefulness, such that users found summaries emphasizing harms to be more useful than those emphasizing health benefits, \textit{F}(1, 1205) = 6.97, \textit{p} = .008, $\eta_{p}^{2}$ = .030. There was no significant main effect of AI summary placement (top vs. middle) (\textit{p} = .871), and no significant interaction between issue stance and AI placement (\textit{p} = .255). 

The results revealed no significant main effects of the AI summary’s issue stance (\textit{p} = .386) or placement (top vs. middle) (\textit{p} = .943) on users’ trust in the AI-generated summaries. The interaction between issue direction and placement was also not significant (\textit{p} = .622). Therefore, neither the issue stance nor its placement significantly influenced users trust in the AI-generated summaries. 

The results showed no significant main effects of issue stance (\( B = 0.09 \), \( p = .635 \)) or placement (\( B = 0.11 \), \( p = .541 \)) of AI summary, and no significant interaction between the two (\( B = -0.14 \), \( p = .593 \)). Thus, neither the issue stance nor the placement of the AI summary had a substantial effect on users’ willingness to keep or disable the AI summarization feature in the future.

\subsection*{Alternative Explanations}
To address the concern that differences in participants’ responses might be driven by unequal exposure or attention to the stimuli (specifically, that the AI-summary conditions contained more texts than the no-AI-summary control), we compared the time participants spent reading the stimuli page across conditions. The analysis revealed no significant differences in reading time between the AI-summary and no-AI-summary conditions (see Appendix section \rom{6} for descriptive statistics and ANOVA results). This indicates that participants devoted comparable levels of attention and effort across conditions, reducing the likelihood that the observed effects are attributable to differential time spent with the materials rather than the presence or absence of AI-generated summaries.

\section*{Discussion}


Our study is among the first to empirically examine the persuasive effects of AI-generated search summaries in the context of publicly debated issues, a setting that reflects how users increasingly encounter AI-generated content during real-world information seeking. Prior research has demonstrated the persuasive power of AI-generated content across contexts, such as propaganda \cite{goldstein_how_2024}, social media campaigns \cite{dash_persuasive_2025}, and personalized messages \cite{matz_potential_2024}. Our study embedded AI-generated summaries within a simulated online search interface, offers some baseline evidence of how AI-integrated search engine can shape public opinions.

Our findings provide robust support for H1a–c, showing that the presence (vs. absence) of an AI-generated summary on a search results page significantly shifted users’ attitudes, behavioral intentions, and policy support toward the stance advocated in the summary. This aligns with prior studies highlighting the persuasive potential of AI-generated content across diverse domains including public health and propaganda \cite{dash_persuasive_2025, matz_potential_2024, goldstein_how_2024}. Users likely relied on cognitive shortcuts when engaging with search results; the AI-generated summary might have simplified complex information, prompting users to adopt the viewpoint it conveyed without deep scrutiny \cite{eagly_psychology_1993, tversky_judgment_1974}. These results suggest that AI-generated search summaries can be used as persuasive communicators. Their presence, even with diverse search results on the same page, can guide public opinions regarding publicly debated or ambiguous topics.

The significant effect of summary position on attitudes (H2a) supports the notion that users anchor on information presented at the top of the page, consistent with established research on anchoring and primacy effects in digital environments \cite{tversky_judgment_1974, brunel_message_2003, bhattacharya_examining_2024}. However, anchoring influenced initial judgments (attitudes; H2a supported) more than downstream behavioral intentions (H2b not supported) or policy support (H2c not supported). This might reflect reduced salience of summary position when users evaluate more consequential outcomes. Although the AI summary placement only significantly influenced attitudes (H2a), the consistent effects in all outcomes for the mere presence of an AI-generated summary (H1a-c) underscores the power of AI-generated content to shape user beliefs even when not prominently placed, raising heightened concerns for platform design.

The persuasive effects of benefit-framed AI summaries were significantly stronger among users who were less familiar with the health issue, particularly for issue attitudes and behavioral intentions, whereas harm-framed summaries were less affected by familiarity and had more consistent effects overall (RQ2a). This pattern aligns with classic findings in persuasion research, which suggest that persuasive messages are more effective when individuals expect or perceive the content to be novel, as unfamiliar arguments are more likely to elicit opinion change \cite{sears_effects_1965}. In our case, users with lower familiarity may have treated the AI-generated summary as introducing new, un-encountered information, increasing their receptivity to its content. This suggests that AI-generated search summaries may exert the greatest persuasive influence when users seek information on less familiar topics, precisely when they are most vulnerable to novel framing cues and most reliant on heuristic processing, amplifying the potential for AI to shape public understanding and decision-making in domains where baseline knowledge is low.

Users with higher general trust in AI were more persuaded by benefit-framed AI summaries (vs. control) in terms of issue attitudes and behavioral intentions, and more persuaded by harm-framed summaries (vs. control) in terms of behavioral intentions and policy support (RQ2b). This reflects the broader literature on source credibility, where greater trust in the source increases message acceptance \cite{hohenberg_when_2023}. The moderating role of general AI trust also highlights potential risks of over-reliance on AI-generated summaries, especially among users who view AI as highly credible. Designers and policymakers should consider transparency and labeling strategies to ensure AI systems do not inadvertently reinforce biased, one-sided, or false information.

One possible explanation for users' greater perceived usefulness (RQ3) is that risk-oriented messages are often seen as more transparent and credible. Prior research suggests that people perceive communicators who acknowledge risks or uncertainties as more honest and trustworthy, compared to those who focus solely on benefits \cite{peters_functions_2006, ahluwalia_consumer_2000, okeefe_how_1999}. This aligns with theories of negativity bias \cite{rozin_negativity_2001}, which suggest that highlighting potential harms may signal objectivity and caution, thereby increasing users’ confidence in the message. This finding also raises important concerns: while risk-aversion framing may increase engagement and reliance, it could also amplify the persuasive power of misleading or exaggerated risk claims, especially in high-stake contexts such as health related issues. This underscores the need for careful design, auditing, and transparency mechanisms when deploying AI systems that summarize sensitive health information.

\section*{Limitations}

One limitation of our study is the use of passive exposure to the static search results page with non-clickable mock search results, which differs from real-world, goal-directed search behaviors with clickable search results. However, this design allowed us to systematically isolate the effects of AI presence, placement, and framing; it also mirrors how users may incidentally encounter AI-generated content during everyday browsing. We also incorporated clickable search results stimuli in one of our subsequent studies (currently under review) to address this limitation. 

A related limitation is that, in real-world contexts, search results vary based on user-generated queries \cite{sharma_generative_2024}, which are often shaped by individuals’ preexisting attitudes or beliefs. These personalized search paths can introduce selection bias and complicate the interpretation of AI content effects. Another important consideration is the role of source cues in how users process search results. While we included mock source links to reduce confounding effects of source credibility, this decision may have lowered the ecological validity of the stimuli. Nonetheless, conducting a preliminary, well-controlled experiment is a critical step in establishing baseline effects before moving to more ecologically complex designs. Future research should incorporate interactive, user-driven search tasks to explore how search motivation, prior attitudes, source credibility, and other factors can influence both query formulation and the processing of AI-generated summaries.

In addition, our study design only reflects current features of AI-integrated online information-seeking tools, but AI tools and users’ familiarity with these tools continue to evolve, posing challenges for long-term generalizability. Nonetheless, our findings identify foundational psychological mechanisms (e.g., heuristic processing) that are likely to persist even as AI interface designs change. By focusing on user cognition, we aim to generate insights applicable across contexts and resilient to evolving technologies. Future studies can build on this by examining how emerging AI interface features or summary styles activate different psychological mechanisms, shaping user judgments across diverse contexts and populations.

\section*{Conclusion}

In conclusion, this study provides some of the first empirical evidence that AI-generated summaries on search result pages can significantly shape users' attitudes, behavioral intentions, and policy support on publicly debated issues. The presence of AI search summaries, even when not prominently placed, acts as a persuasive cue, especially for users with lower topic familiarity or higher trust in AI. Harm-framed (vs. benefit-framed) AI summary was seen as more useful, raising concerns about risk amplification. These findings underscore the need for transparent, accountable AI design and regulation in information systems, and point to future directions for understanding AI influence in more interactive, goal-directed search contexts.

\section*{Methods and Materials}

This study was approved by the University of Washington Institutional Review Board (IRB), and the experiment was performed in accordance with relevant guidelines and regulations. 

\subsection*{Study Procedure \& Experimental Design}

The study employed a [2 AI summary positions × 2 AI summary stances + 1 no-AI-summary control] × 4 issues design, resulting in 20 experimental conditions. The AI summary stance and issue factors were included to enhance the generalizability of the findings, but they were not central to our preregistered hypotheses. The main analysis does not focus on differences across stances or across the four health issues. 
Specifically, we manipulated the presence (versus absence) and the placement (top versus mid) of the AI-generated summary. The stance of the AI summary includes emphasizing benefits or harms. This design results in five analytic conditions across four different issues: (1) Top-benefit (\textit{n} = 406): AI summary emphasizing benefits appears at the top of the search page with randomized search results; (2) Mid-benefit (\textit{n} = 401): Benefit summary appears in the middle of the search page with randomized search results; (3) Top-harm (\textit{n} = 402): AI summary emphasizing harms appears at the top of the search page with randomized search results; (4) Mid-harm (\textit{n} = 399): Harm summary appears in the middle of the search page with randomized search results; (5) Control (\textit{n} = 396): No AI-generated summary included, but the search page still includes the same mock search results as other conditions. 
We selected four health topics: raw milk (\textit{n} = 498), water fluoridation (\textit{n} = 498), artificial sweeteners (\textit{n} = 505), and GMO foods (\textit{n} = 503). Participants were randomly assigned to one of these topics to minimize item-specific effects and enhance ecological validity. Our topic selection was informed in part by the KFF Health (Mis)information Monitor\footnote{\url{https://www.kff.org/tag/the-monitor/}}, which tracks emerging and contested health issues. These topics were well-suited for our study because of their relevance to ongoing public debates and their potential to elicit diverse claims and perspectives, thereby better approximating real-world conditions in which users may encounter AI-generated summaries on debated health topics.

\begin{figure} 
	\centering
	\includegraphics[width=0.95\textwidth]{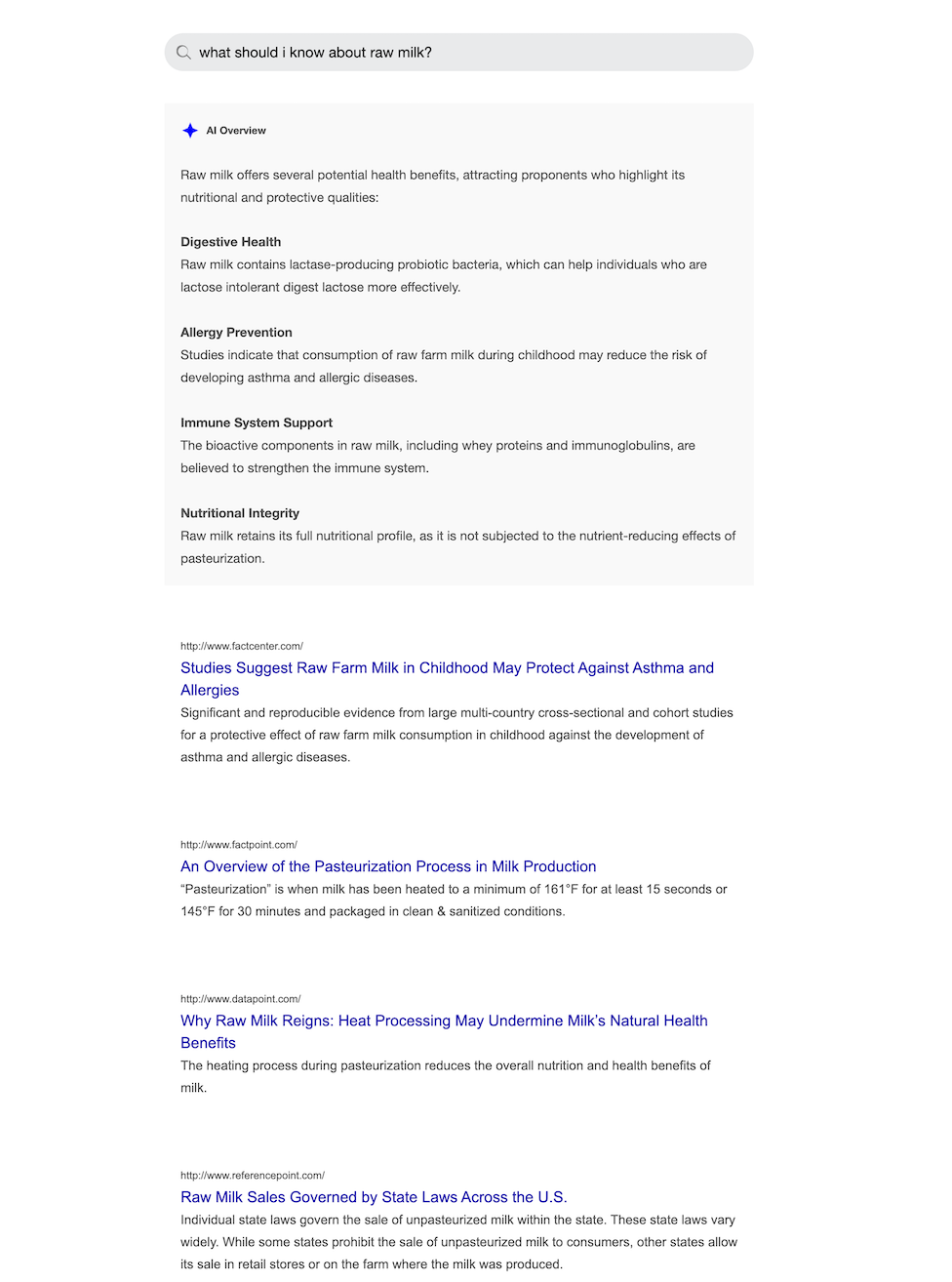} 

	\caption{\textbf{Sample stimuli.} Benefit-framed AI summary (raw milk) placed at the top.
        }
	\label{fig:stimuli} 
\end{figure}

Participants who consented to the study were randomly assigned to one of the five experimental conditions among one of the four issues. Participants were first instructed to browse the mock search interface simulating a standard search engine results page (See Figure \ref{fig:stimuli} for a sample stimuli about raw milk). Following this exposure, participants completed a survey questionnaire (see Measures), concluded with demographic questions, and a debriefing when their assigned stimuli contained factually wrong information. 

\subsubsection*{Participants Demographics and Randomization Checks}

We used Prolific (an online crowdsourcing platform) and recruited a sample of U.S.-based adults (\textit{N} = 2,004, April 2025), representative of age, sex, and political affiliation, using Prolific's prescreening criteria\footnote{\url{https://researcher-help.prolific.com/en/article/e6555f}}. A priori power analysis  determined that n = 400 in each condition would detect effect size (Cohen's d) $\geq$ 0.20 at $\alpha$ = .05 (adjusted for multiple comparisons) with power = .8. Informed consent was obtained from all subjects. We paid participants \$1.00 (hourly rate: \$13.14) for completing the study, the median completion time was 4 minutes and 34 seconds. 
The demographic composition of participants is shown in Table S1 in the Appendix. The randomization check, also reported in Table S1, indicated that all measured demographic variables were balanced across conditions; therefore, none were included as covariates in the analyses.

\subsubsection*{Mock Search Results Generation and Validation}

In the stimuli, each mock result included a headline, short overview, and fictitious URL, designed to mirror typical content found in online searches (See Figure \ref{fig:stimuli}). We first selected information about the benefits and harms of four controversial health issues -- raw milk, water fluoridation, artificial sweeteners, and genetically modified organisms (GMOs) -- from credible sources, including the CDC and FDA (see Appendix section \rom{1} for the full list of mock search results and content sources). Based on this information, we created titles and overviews that resembled authentic online search results. For each topic, we generated 12 search results: four emphasizing health benefits, four emphasizing harms, and four presenting neutral information (e.g., providing a definition of raw milk). To enhance the visual ecological validity of the mock search interface, each search result was paired with a fictitious source link. We opted not to use real URLs to prevent source credibility from acting as a confounding variable. The twelve mock results were presented in randomized order to control for potential ordering effects.

To ensure each result accurately reflected the intended stance (i.e., emphasizing benefit, emphasizing harm, neutral), we used \texttt{o1-2024-12-17} to annotate the perceived stance of the headlines for a given topic (refer to Appendix section \rom{1} for details on the o1-annotator).

\subsubsection*{AI Summary Generation}

To generate an AI summary of the mock search engine results page, we explicitly instructed \texttt{gpt-4o} to generate a 100-word summary of the search results collected above. We specified in our prompt (see Appendix section \rom{1} for the exact prompt) to only emphasize either the harms or the benefits of the given health topic. A full list of AI summaries we generated can be found in Appendix section \rom{1}.
The AI summaries were generated solely from the headlines and snippets contained in the mock search results and did not introduce any additional information beyond what appeared in the control condition. Thus, experimental conditions with AI summaries did not provide participants with more substantive information than the control; they differed only in the inclusion of an AI summary.

\subsubsection*{Manipulation Checks}
Because our experimental manipulations operated entirely through the presence, position, and stance of the AI-generated summaries, the summary-generation process itself serves as the manipulation validation. We produced all summaries in advance using a controlled prompt that required \texttt{gpt-4o} to generate a 100-word summary based on the same set of search results and to adopt a clearly defined benefit-focused or harm-focused stance. We then verified that each summary reflected the intended framing and matched the designated placement (top or middle) in the search interface. Since the stimuli were fixed, pre-validated, the manipulation validity is assured at the stimulus level without requiring an additional participant-level check.

\subsection*{Measures}

A full list of preregistered measurement items and scales can be found at OSF: 
\url{https://osf.io/rsyhg?view_only=834769ba56094c3cbbc236f353e301c0}.

\paragraph{Issue Attitudes.}
We measured participants’ attitudes toward the targeted health issue (i.e., drinking raw milk; adding fluoride to local water supplies; adding artificial sweeteners to one’s diet; eating GMO foods) using three 6-point bipolar scales: \textit{bad–good}, \textit{harmful–beneficial}, and \textit{foolish–wise}. These items were adapted from the Theory of Planned Behavior (TPB) \cite{ajzen_theory_1991}. Responses were averaged from the three statements to form a composite attitude score, with higher values indicating more favorable attitudes (Cronbach’s $\alpha$ = 0.97).

\paragraph{Behavioral Intentions.}
Participants reported how likely they were to engage in the health behavior in the future (e.g., drinking raw milk; consuming artificial sweeteners). Each intention was assessed on a 6-point Likert scale ranging from very unlikely to very likely. Items were adapted from TPB \cite{ajzen_theory_1991} and treated as single-item indicator.

\paragraph{Policy Support.}
Support for restrictive policies was measured using one policy statement for each topic (e.g., ``The sale of unpasteurized or raw milk should be restricted''). Participants rated their support on a 6-point scale from strongly oppose to strongly support. These items were collected from reputable sources (e.g., Associated Press, FDA, etc.) and modeled after prior research on public support for health and science policies \cite{niederdeppe_evidence-based_2021}.

\paragraph{Issue Familiarity.}
Participants retrospectively rated their familiarity with the assigned topic in terms of its health-related effects, using a 6-point scale ranging from not familiar to very familiar. This measure was administered after stimulus exposure to avoid priming participants to focus excessively on the target topic in advance.

\paragraph{AI Summary Use and Trust.}
Participants answered a set of items regarding their experience with the AI-generated search summary. \textit{Perceived AI presence} was measured with a dichotomous item (``Did the search results you saw earlier contain an AI Summary?''; yes/no). Those who indicated ``yes'' were asked whether they read or skipped the summary. If they read it, they rated its usefulness on a 6-point scale from very useless to very useful. Participants then reported their trust in the AI summary they viewed to provide accurate information, using a 6-point Likert scale ranging from not at all to very much. Participants also indicated whether they would (or would not) disable the AI summarizing function if they had the option to do so.

\paragraph{General Trust in AI.}
General trust in AI was measured by asking participants how much they trust AI to provide accurate information most of the time, using a 6-point Likert scale ranging from not at all to very much. 

\subsection*{Hypotheses and Research Questions}

Preregistered Hypothesis 1a: The presence (vs. absence) of an AI-generated summary influences user beliefs by moving their \textit{issue attitudes} closer to the stance advocated in the summary.
\\ 
Preregistered Hypothesis 1b: The presence (vs. absence) of an AI-generated summary influences user beliefs by moving their \textit{behavioral intentions} closer to the stance advocated in the summary.
\\ 
Preregistered Hypothesis 1c: The presence (vs. absence) of an AI-generated summary influences user beliefs by moving their \textit{policy support} closer to the stance advocated in the summary.
\\ \\
Preregistered Hypothesis 2a: When the AI-generated summary is present, its position on the page (top vs. middle) influences user beliefs through anchoring, with top positioning exerting a stronger persuasive effect by moving their \textit{issue attitudes} closer to the stance advocated in the summary.
\\ 
Preregistered Hypothesis 2b: When the AI-generated summary is present, its position on the page (top vs. middle) influences user beliefs through anchoring, with top positioning exerting a stronger persuasive effect by moving their \textit{behavioral intentions} closer to the stance advocated in the summary.
\\ 
Preregistered Hypothesis 2c: When the AI-generated summary is present, its position on the page (top vs. middle) influences user beliefs through anchoring, with top positioning exerting a stronger persuasive effect by moving their \textit{policy support} closer to the stance advocated in the summary.
\\ \\ 
Exploratory Research Question 1: Will the issue stance of AI-generated summary moderate its effects?
\\ \\ 
Exploratory Research Question 2a: Will users' issue familiarity and (b) general trust in AI moderate the effects of AI-generated summary?
\\ \\ 
Exploratory Research Question 2b: Will users' general trust in AI moderate the effects of AI-generated summary?
\\ \\ 
Exploratory Research Question 3: How will AI-generated summary differing in its issue direction and position influence users’ perception of such summary, including trustworthiness and usefulness?

\subsection*{Analytical Approach}

To answer H1, we first fitted a one‐way ANCOVA, specifying the three‐level independent factor (benefit-framed AI summaries, harm-framed AI summaries, control) as the independent variable and including topic, user issue familiarity, and user AI trust as covariates. The test of H1 was conducted using a joint test of two planned contrasts -- benefit-framed AI vs. control, harm-framed AI vs. control. In addition, we conducted pairwise comparison between these two planned contrasts to test whether the effect of the presence (versus absence) of AI summary was moderated by issue stance (RQ1). We repeated all above these steps for each dependent variable: (a) issue attitudes, (b) behavioral intentions, and (c) policy support.

To answer H2, we first fitted a one‐way ANCOVA, specifying the five‐level independent factor (benefit-framed AI summary at top, benefit-framed AI summary at mid, harm-framed AI summary at top, harm-framed AI summary at mid, control) as the independent variable and including topic, user issue familiarity, and user AI trust as covariates. The test of H2 was conducted using a joint test of two planned contrasts -- benefit-framed AI summary at top vs. benefit-framed AI summary at mid, harm-framed AI summary at top vs. harm-framed AI summary at middle. In addition, we conducted pairwise comparison between these two planned contrasts to test whether the effect of the placement (top vs. mid) of AI summary was moderated by issue stance (RQ1). We repeated all above these steps for each dependent variable: (a) issue attitudes, (b) behavioral intentions, and (c) policy support.

The analyses for H1 and H2 deviated slightly from the preregistered plan to better align with our theoretical hypotheses and improve statistical efficiency. Detailed justifications for these deviations are provided in Appendix section \rom{4}.


To address RQ2a about the moderating role of users' issue familiarity, we conducted OLS regression models in which we included the dummy coded condition variables (with control condition as the reference group to test the effects of AI summary presence; with mid-placed AI summary condition as the reference group to test the effects of AI summary placement), the moderator (issue familiarity), and interaction terms between the targeted condition comparison and the moderator to predict each of the outcome variables. When the interaction term between condition and moderator was (marginally) statistically significant (as recommended by D’Alonzo and colleagues \cite{dalonzo_johnson-neyman_2004}), we used Johnson-Neyman technique to identify the specific values of moderator at which the effect of condition on the outcome became significant. We repeated all above these steps for each dependent variable: (a) issue attitudes, (b) behavioral intentions, and (c) policy support.
To address RQ2b about the moderating role of users' general trust in AI, we repeated the same steps by including trust as the moderator.

To address RQ3, how AI-generated summaries influence user experience, we examined the effects of summary stance and placement on three key dependent variables: perceived usefulness and trustworthiness of the AI-generated summary, as well as users' intention in keeping or disabling AI summarization function in the future. 
For the continuous outcome variables, we conducted two sets of 2 (issue stance: benefit vs. harm) × 2 ANCOVAs, one for each dependent variable (perceived usefulness and trustworthiness ). The first ANCOVA tested effects on perceived usefulness of the AI summary, and the second tested effects on trustworthiness. In both models, we included topic, familiarity, and trust as covariates. Each model was tested for main effects of summary issue stance (whether the AI summary emphasized benefits vs. harms) and main effects of summary placement (whether the summary appeared at the top vs. in the middle of the search results), and the interaction effect between issue stance and placement. We applied Tukey-adjusted post hoc contrasts to further probe significant main and interaction effects. 
For the binary outcome variable, we conducted a logistic regression predicting the outcome (will disable versus will not disable) using issue stance (benefit- vs. harm-framed), placement (top vs. mid), their interaction, and covariates including topic, issue familiarity, and general trust in AI. 
All analyses for this section were restricted to non-control conditions.

We employed robust standard errors (SE) for all analyses reported above to account for potential heteroskedasticity.

\subsection*{Data Availability}
The datasets used and/or analyzed during the current study are available from the corresponding author on reasonable request. Datasets and analytic codes will be published upon the revision and publication of this paper.

\bibliography{zotero-references}

@misc{eddy_americans_2025,
	title = {Americans have mixed feelings about {AI} summaries in search results},
	url = {https://www.pewresearch.org/short-reads/2025/10/01/americans-have-mixed-feelings-about-ai-summaries-in-search-results/},
	language = {en-US},
	urldate = {2025-11-27},
	journal = {Pew Research Center},
	author = {Eddy, Kirsten},
	month = oct,
	year = {2025},
}

@article{dash_persuasive_2025,
	title = {The persuasive potential of {AI}-paraphrased information at scale},
	volume = {4},
	issn = {2752-6542},
	url = {https://doi.org/10.1093/pnasnexus/pgaf207},
	doi = {10.1093/pnasnexus/pgaf207},
	number = {7},
	urldate = {2025-09-10},
	journal = {PNAS Nexus},
	author = {Dash, Saloni and Xu, Yiwei and Jalbert, Madeline and Spiro, Emma S},
	month = jul,
	year = {2025},
	pages = {pgaf207},
}

@article{dalonzo_johnson-neyman_2004,
	title = {The {Johnson}-{Neyman} procedure as an alternative to {ANCOVA}},
	volume = {26},
	issn = {0193-9459},
	doi = {10.1177/0193945904266733},
	language = {eng},
	number = {7},
	journal = {Western Journal of Nursing Research},
	author = {D'Alonzo, Karen T.},
	month = nov,
	year = {2004},
	pmid = {15466616},
	pmcid = {PMC3203541},
	pages = {804--812},
}

@article{matz_potential_2024,
	title = {The potential of generative {AI} for personalized persuasion at scale},
	volume = {14},
	copyright = {2024 The Author(s)},
	issn = {2045-2322},
	url = {https://www.nature.com/articles/s41598-024-53755-0},
	doi = {10.1038/s41598-024-53755-0},
	language = {en},
	number = {1},
	urldate = {2025-05-29},
	journal = {Scientific Reports},
	author = {Matz, S. C. and Teeny, J. D. and Vaid, S. S. and Peters, H. and Harari, G. M. and Cerf, M.},
	month = feb,
	year = {2024},
	note = {Publisher: Nature Publishing Group},
	pages = {4692},
}

@misc{williams-ceci_biased_2025,
	title = {Biased {AI} writing assistants shift users’ attitudes on societal issues},
	url = {https://osf.io/mhjn6_v2},
	doi = {10.31234/osf.io/mhjn6_v2},
	language = {en-us},
	urldate = {2025-07-18},
	publisher = {OSF},
	author = {Williams-Ceci, Sterling and Jakesch, Maurice and Bhat, Advait and Kadoma, Kowe and Zalmanson, Lior and Naaman, Mor},
	month = feb,
	year = {2025},
}

@article{mieczkowski_ai-mediated_2021,
	title = {{AI}-mediated communication: {Language} use and interpersonal effects in a referential communication task},
	volume = {5},
	shorttitle = {{AI}-{Mediated} {Communication}},
	url = {https://dl.acm.org/doi/10.1145/3449091},
	doi = {10.1145/3449091},
	number = {CSCW1},
	urldate = {2025-07-18},
	journal = {Proc. ACM Hum.-Comput. Interact.},
	author = {Mieczkowski, Hannah and Hancock, Jeffrey T. and Naaman, Mor and Jung, Malte and Hohenstein, Jess},
	month = apr,
	year = {2021},
	pages = {17:1--17:14},
}

@article{karinshak_working_2023,
	title = {Working with {AI} to persuade: {Examining} a large language model's ability to generate pro-vaccination messages},
	volume = {7},
	shorttitle = {Working {With} {AI} to {Persuade}},
	url = {https://dl.acm.org/doi/10.1145/3579592},
	doi = {10.1145/3579592},
	number = {CSCW1},
	urldate = {2025-05-29},
	journal = {Proc. ACM Hum.-Comput. Interact.},
	author = {Karinshak, Elise and Liu, Sunny Xun and Park, Joon Sung and Hancock, Jeffrey T.},
	month = apr,
	year = {2023},
	pages = {116:1--116:29},
}

@article{niederdeppe_evidence-based_2021,
	title = {Evidence-{Based} {Message} {Strategies} to increase public support for state investment in early childhood education: {Results} from a longitudinal panel experiment},
	volume = {99},
	copyright = {© 2021 The Authors. The Milbank Quarterly published by Wiley Periodicals LLC on behalf of The Milbank Memorial Fund},
	issn = {1468-0009},
	shorttitle = {Evidence-{Based} {Message} {Strategies} to {Increase} {Public} {Support} for {State} {Investment} in {Early} {Childhood} {Education}},
	url = {https://onlinelibrary.wiley.com/doi/abs/10.1111/1468-0009.12534},
	doi = {10.1111/1468-0009.12534},
	language = {en},
	number = {4},
	urldate = {2025-06-23},
	journal = {The Milbank Quarterly},
	author = {Niederdeppe, Jeff and Winett, Liana B. and Xu, Yiwei and Fowler, Erika Franklin and Gollust, Sarah E.},
	year = {2021},
	note = {\_eprint: https://onlinelibrary.wiley.com/doi/pdf/10.1111/1468-0009.12534},
	pages = {1088--1131},
}

@article{ajzen_theory_1991,
	title = {The theory of planned behavior},
	volume = {50},
	issn = {1095-9920},
	doi = {10.1016/0749-5978(91)90020-T},
	number = {2},
	journal = {Organizational Behavior and Human Decision Processes},
	author = {Ajzen, Icek},
	year = {1991},
	note = {Place: Netherlands
Publisher: Elsevier Science},
	pages = {179--211},
}

@article{rothman_shaping_1997,
	title = {Shaping perceptions to motivate healthy behavior: {The} role of message framing},
	volume = {121},
	issn = {1939-1455},
	shorttitle = {Shaping perceptions to motivate healthy behavior},
	doi = {10.1037/0033-2909.121.1.3},
	number = {1},
	journal = {Psychological Bulletin},
	author = {Rothman, Alexander J. and Salovey, Peter},
	year = {1997},
	note = {Place: US
Publisher: American Psychological Association},
	pages = {3--19},
}

@article{okeefe_relative_2007,
	title = {The relative persuasiveness of gain-framed loss-framed messages for encouraging disease prevention behaviors: {A} meta-analytic review},
	volume = {12},
	issn = {1081-0730},
	shorttitle = {The {Relative} {Persuasiveness} of {Gain}-{Framed} {Loss}-{Framed} {Messages} for {Encouraging} {Disease} {Prevention} {Behaviors}},
	url = {https://doi.org/10.1080/10810730701615198},
	doi = {10.1080/10810730701615198},
	number = {7},
	urldate = {2025-06-10},
	journal = {Journal of Health Communication},
	author = {O'Keefe, Daniel J. and and Jensen, Jakob D.},
	month = oct,
	year = {2007},
	pmid = {17934940},
	note = {Publisher: Taylor \& Francis
\_eprint: https://doi.org/10.1080/10810730701615198},
	pages = {623--644},
}

@article{gallagher_health_2012,
	title = {Health message framing effects on attitudes, intentions, and behavior: {A} meta-analytic review},
	volume = {43},
	issn = {0883-6612},
	shorttitle = {Health {Message} {Framing} {Effects} on {Attitudes}, {Intentions}, and {Behavior}},
	url = {https://doi.org/10.1007/s12160-011-9308-7},
	doi = {10.1007/s12160-011-9308-7},
	number = {1},
	urldate = {2025-06-10},
	journal = {Annals of Behavioral Medicine},
	author = {Gallagher, Kristel M. and Updegraff, John A.},
	month = feb,
	year = {2012},
	pages = {101--116},
}

@article{reder_comparison_1980,
	title = {A comparison of texts and their summaries: {Memorial} consequences},
	volume = {19},
	issn = {0022-5371},
	shorttitle = {A comparison of texts and their summaries},
	url = {https://www.sciencedirect.com/science/article/pii/S002253718090122X},
	doi = {10.1016/S0022-5371(80)90122-X},
	number = {2},
	urldate = {2025-06-10},
	journal = {Journal of Verbal Learning and Verbal Behavior},
	author = {Reder, Lynne M. and Anderson, John R.},
	month = apr,
	year = {1980},
	pages = {121--134},
}

@article{rozin_negativity_2001,
	title = {Negativity bias, negativity dominance, and contagion},
	volume = {5},
	issn = {1088-8683},
	url = {https://doi.org/10.1207/S15327957PSPR0504_2},
	doi = {10.1207/S15327957PSPR0504_2},
	language = {EN},
	number = {4},
	urldate = {2025-06-06},
	journal = {Personality and Social Psychology Review},
	author = {Rozin, Paul and Royzman, Edward B.},
	month = nov,
	year = {2001},
	note = {Publisher: SAGE Publications Inc},
	pages = {296--320},
}

@article{ahluwalia_consumer_2000,
	title = {Consumer response to negative publicity: {The} moderating role of commitment},
	volume = {37},
	issn = {0022-2437},
	shorttitle = {Consumer {Response} to {Negative} {Publicity}},
	url = {https://doi.org/10.1509/jmkr.37.2.203.18734},
	doi = {10.1509/jmkr.37.2.203.18734},
	language = {EN},
	number = {2},
	urldate = {2025-06-06},
	journal = {Journal of Marketing Research},
	author = {Ahluwalia, Rohini and Burnkrant, Robert E. and Unnava, H. Rao},
	month = may,
	year = {2000},
	note = {Publisher: SAGE Publications Inc},
	pages = {203--214},
}

@article{okeefe_how_1999,
	title = {How to handle opposing arguments in persuasive messages: {A} meta-analytic review of the effects of one-sided and two-sided messages},
	volume = {22},
	issn = {2380-8985},
	shorttitle = {How to {Handle} {Opposing} {Arguments} in {Persuasive} {Messages}},
	url = {https://doi.org/10.1080/23808985.1999.11678963},
	doi = {10.1080/23808985.1999.11678963},
	number = {1},
	urldate = {2025-06-06},
	journal = {Annals of the International Communication Association},
	author = {O’Keefe, Daniel J.},
	month = jan,
	year = {1999},
	note = {Publisher: Routledge
\_eprint: https://doi.org/10.1080/23808985.1999.11678963},
	pages = {209--249},
}

@article{peters_functions_2006,
	title = {The functions of affect in health communications and in the construction of health preferences},
	volume = {56},
	issn = {1460-2466},
	url = {https://onlinelibrary.wiley.com/doi/abs/10.1111/j.1460-2466.2006.00287.x},
	doi = {10.1111/j.1460-2466.2006.00287.x},
	language = {en},
	number = {s1},
	urldate = {2025-06-06},
	journal = {Journal of Communication},
	author = {Peters, Ellen and Lipkus, Isaac and Diefenbach, Michael A.},
	year = {2006},
	note = {\_eprint: https://onlinelibrary.wiley.com/doi/pdf/10.1111/j.1460-2466.2006.00287.x},
	pages = {S140--S162},
}

@article{sears_effects_1965,
	title = {Effects of expected familiarity with arguments upon opinion change and selective exposure},
	volume = {2},
	issn = {1939-1315},
	doi = {10.1037/h0022380},
	number = {3},
	journal = {Journal of Personality and Social Psychology},
	author = {Sears, David W. and Freedman, Jonathan L.},
	year = {1965},
	note = {Place: US
Publisher: American Psychological Association},
	pages = {420--426},
}

@article{hohenberg_when_2023,
	title = {When do sources persuade? {The} effect of source credibility on opinion change},
	volume = {10},
	issn = {2052-2630, 2052-2649},
	shorttitle = {When {Do} {Sources} {Persuade}?},
	url = {https://www.cambridge.org/core/journals/journal-of-experimental-political-science/article/when-do-sources-persuade-the-effect-of-source-credibility-on-opinion-change/48ECC9B706B2C3BA733936BE184917CC},
	doi = {10.1017/XPS.2022.2},
	language = {en},
	number = {3},
	urldate = {2025-06-05},
	journal = {Journal of Experimental Political Science},
	author = {Hohenberg, Bernhard Clemm von and Guess, Andrew M.},
	month = nov,
	year = {2023},
	pages = {328--342},
}

@article{goldstein_how_2024,
	title = {How persuasive is {AI}-generated propaganda?},
	volume = {3},
	issn = {2752-6542},
	url = {https://doi.org/10.1093/pnasnexus/pgae034},
	doi = {10.1093/pnasnexus/pgae034},
	number = {2},
	urldate = {2025-05-29},
	journal = {PNAS Nexus},
	author = {Goldstein, Josh A and Chao, Jason and Grossman, Shelby and Stamos, Alex and Tomz, Michael},
	month = feb,
	year = {2024},
	pages = {pgae034},
}

@book{eagly_psychology_1993,
	address = {Orlando, FL, US},
	series = {The psychology of attitudes},
	title = {The psychology of attitudes},
	isbn = {978-0-15-500097-1},
	publisher = {Harcourt Brace Jovanovich College Publishers},
	author = {Eagly, Alice H. and Chaiken, Shelly},
	year = {1993},
	note = {Pages: xxii, 794},
}

@article{brunel_message_2003,
	title = {Message {Order} {Effects} and {Gender} {Differences} in {Advertising} {Persuasion}},
	volume = {43},
	issn = {1740-1909},
	doi = {10.1017/S0021849903030320},
	number = {3},
	journal = {Journal of Advertising Research},
	author = {Brunel, Frédéric F. and Nelson, Michelle R.},
	year = {2003},
	note = {Place: United Kingdom
Publisher: Cambridge University Press},
	pages = {330--341},
}

@article{tversky_judgment_1974,
	title = {Judgment under uncertainty: {Heuristics} and biases},
	volume = {185},
	issn = {1095-9203},
	shorttitle = {Judgment under uncertainty},
	doi = {10.1126/science.185.4157.1124},
	number = {4157},
	journal = {Science},
	author = {Tversky, Amos and Kahneman, Daniel},
	year = {1974},
	note = {Place: US
Publisher: American Assn for the Advancement of Science},
	pages = {1124--1131},
}

@article{bhattacharya_examining_2024,
	title = {Examining the impact of sponsored search results on choice: {An} anchoring perspective},
	volume = {59},
	issn = {1094-9968},
	shorttitle = {Examining the {Impact} of {Sponsored} {Search} {Results} on {Choice}},
	url = {https://doi.org/10.1177/10949968231195856},
	doi = {10.1177/10949968231195856},
	language = {en},
	number = {2},
	urldate = {2025-02-20},
	journal = {Journal of Interactive Marketing},
	author = {Bhattacharya, Abhi and Robinson, Stacey and Mookherjee, Satadruta and Blöte, Herman},
	month = may,
	year = {2024},
	note = {Publisher: SAGE Publications},
	pages = {183--200},
}

@inproceedings{sharma_generative_2024,
	address = {New York, NY, USA},
	series = {{CHI} '24},
	title = {Generative {Echo} {Chamber}? {Effect} of {LLM}-{Powered} {Search} {Systems} on {Diverse} {Information} {Seeking}},
	isbn = {9798400703300},
	shorttitle = {Generative {Echo} {Chamber}?},
	url = {https://dl.acm.org/doi/10.1145/3613904.3642459},
	doi = {10.1145/3613904.3642459},
	urldate = {2025-02-19},
	booktitle = {Proceedings of the 2024 {CHI} {Conference} on {Human} {Factors} in {Computing} {Systems}},
	publisher = {Association for Computing Machinery},
	author = {Sharma, Nikhil and Liao, Q. Vera and Xiao, Ziang},
	month = may,
	year = {2024},
	pages = {1--17},
}

@misc{memon_search_2024,
	title = {Search {Engines} {Post}-{ChatGPT}: {How} {Generative} {Artificial} {Intelligence} {Could} {Make} {Search} {Less} {Reliable}},
	shorttitle = {Search {Engines} {Post}-{ChatGPT}},
	url = {http://arxiv.org/abs/2402.11707},
	doi = {10.48550/arXiv.2402.11707},
	urldate = {2025-02-20},
	publisher = {arXiv},
	author = {Memon, Shahan Ali and West, Jevin D.},
	month = feb,
	year = {2024},
	note = {arXiv:2402.11707 [cs]},
}


\section*{Acknowledgments}
We thank Drs. Xinyi Zhou and Amy X. Zhang for their feedback and edits on the initial research proposal.

\section*{Funding}
We thank University of Washington Population Health Initiative (PHI) Tier 1 Research Grant for the funding support. 

\section*{Author contributions statement}


Y.~X. and S.~D. conceived the project and designed the experiment; Y.~X., S.~D., S.~K. constructed the study materials and collected data; Y.~X. led the data analyses, results interpretation, and manuscript writing, W.~L., S.~D., S.~K. supported the analyses, results interpretation, and manuscript writing; E.~S.~S. supervised the project and edited the manuscript.

\section*{Additional information}

\textbf{Competing interests} There are no competing interests to declare. 
\\
\textbf{Supplementary materials} Supplementary materials are available to download. 




\setcounter{table}{0} 
\renewcommand{\thetable}{S\arabic{table}} 

\setcounter{figure}{0} 
\renewcommand{\thefigure}{S\arabic{figure}} 

\begin{center}
\section*{\centering Supplementary Materials}

	Yiwei~Xu$^{\ast}$,
	Saloni~Dash,
	Sungha~Kang,
    Wang~Liao,
    Emma~S.~Spiro\\
\small$^\ast$Corresponding author. Email: yiweixu@umd.edu\and
\end{center}

\tableofcontents

\renewcommand{\thesection}{\Roman{section}}

\clearpage
\section{Stimuli Construction and Validation}
\label{appendix:stimuli}

\subsection*{Mock Search Results Headlines and Overviews}
 
The full list of content sources for all mock search results can be found at OSF: 
\url{https://osf.io/nj5w3?view_only=10824493bbd745799eaf6c5c2d85c137}

\subsubsection*{Topic: Raw milk}

\textbf{Stance: Benefits of raw milk}

(1) \textit{Raw Milk’s Natural Advantage: How Probiotic Bacteria May Ease Lactose Intolerance.} Raw milk does not cause lactose intolerance because it contains lactase secreted by probiotic bacteria present in raw milk.

(2) \textit{Studies Suggest Raw Farm Milk in Childhood May Protect Against Asthma and Allergies.} Evidence from large studies indicates raw farm milk consumption during childhood may reduce asthma and allergy development.

(3) \textit{Raw Farm Milk: A Complex Substance Packed with Immune-Boosting Components.} Raw milk enhances immunity and includes bioactive components like whey proteins, transcription factors, and omega-3 fats.

(4) \textit{Why Raw Milk Reigns: Heat Processing May Undermine Milk’s Natural Health Benefits.} Pasteurization reduces nutritional and health benefits inherent in raw milk.

\textbf{Stance: Harms of raw milk}

(5) \textit{Raw Milk Consumption Poses Health Risks, Especially for Vulnerable Populations.} Raw milk can contain pathogens like E. coli and Salmonella, posing significant health risks.

(6) \textit{Bird Flu Found in Raw Milk: A Dangerous New Twist in Dairy.} Raw milk from cows infected with bird flu (H5N1 virus) poses health threats.

(7) \textit{Raw Milk Linked to Severe Illnesses, Including Kidney Failure and Paralysis.} Bacteria in raw milk can cause life-threatening diseases like Guillain-Barré syndrome.

(8) \textit{Raw Milk Hype Ignores a Century of Proven Safety in Pasteurization.} Pasteurized milk offers nutritional benefits safely, without raw milk's health risks.

\textbf{Neutral Reference}

(9) \textit{Raw Milk Defined: Understanding Its Production and Pasteurization Process.} Raw milk is milk that has not been pasteurized.

(10) \textit{Raw Milk Sales Governed by State Laws Across the U.S.} State-specific regulations control raw milk sales.

(11) \textit{What Is Raw Milk? Unpasteurized Milk from Cows, Goats, Sheep, and More.} Raw milk is unpasteurized milk from various animals, including cows and goats.

(12) \textit{An Overview of the Pasteurization Process in Milk Production.} Pasteurization involves heating milk to kill harmful bacteria.

\subsubsection*{Topic: Water fluoridation}

\textbf{Stance: Benefits of water fluoridation}

(1) \textit{How Water Fluoridation Reduces Cavities and Strengthens Teeth.} Water fluoridation reduces tooth decay by providing frequent and consistent contact with low levels of fluoride, reducing cavities by about 25

(2) \textit{Fluoridated Water Linked to Fewer Cavities in School Children.} Schoolchildren in fluoridated communities have, on average, 2.25 fewer decayed teeth compared to those in non-fluoridated communities.

(3) \textit{Cost-Effective Prevention: Fluoridated Water Reduces Dental Expenses and Health Inequalities.} Fluoridated water provides a return on investment of \$20 for every \$1 spent, saving approximately \$32 per person annually in dental expenses.

(4) \textit{How Water Fluoridation Protects Oral Health Across All Ages.} Fluoridation benefits entire communities by preventing cavities, reducing oral health disparities, and lowering overall dental treatment costs.

\textbf{Stance: Harms of water fluoridation}

(5) \textit{Fluoride in Tap Water Linked to Enamel Damage in Kids' Developing Teeth.} High fluoride levels during childhood can cause dental fluorosis, marked by tiny white streaks or specks on tooth enamel.

(6) \textit{Hidden Cost of Fluoridation: Bone Loss and Hormone Imbalance Tied to Fluoride Exposure.} Excess fluoride exposure can lead to hyperparathyroidism, causing calcium depletion in bones, increasing fracture risk.

(7) \textit{Hidden Danger in Your Tap: Long-term Fluoride Exposure Can Cause Skeletal Damage and Joint Pain.} Prolonged fluoride exposure can cause skeletal fluorosis, resulting in hardened bones, joint pain, and impaired mobility.

(8) \textit{Prenatal Fluoride Exposure from Tap Water May Harm Children’s Cognitive Development.} Higher prenatal fluoride exposure has been linked to lower IQ scores and poorer cognitive outcomes in children.

\textbf{Neutral Reference}

(9) \textit{Fluorides Explained: Natural Occurrence, Uses, and Effects.} Fluorides combine fluorine with other substances like sodium fluoride, stannous fluoride, or MFP fluoride.

(10) \textit{The Applications of Fluoride in Daily Life.} Fluoride occurs naturally and is used synthetically in drinking water, toothpaste, mouthwash, and other products.

(11) \textit{How Communities Implement Water Fluoridation.} Decisions to fluoridate water are typically made by state or local governments, sometimes involving community votes.

(12) \textit{Community Water Fluoridation and Fluoride Levels.} Fluoridation involves adjusting fluoride levels in drinking water according to public health guidelines.

\subsubsection*{Topic: Artificial sweeteners}

\textbf{Stance: Benefits of artificial sweeteners}

(1) \textit{Artificial sweeteners are safe in limited amounts for healthy people.} In general, artificial sweeteners are safe in limited amounts for healthy individuals, including pregnant people. When used in moderation, they can be part of a healthy diet.

(2) \textit{Sweet Without the Guilt: Artificial Sweeteners May Aid Short-Term Weight Management.} Sugar substitutes, being low-calorie or calorie-free, might help adults and children manage weight in the short term.

(3) \textit{Artificial Sweeteners and Oral Health: How Sugar Substitutes Reduce Cavity Risk.} Replacing added sugars with sugar substitutes can lower the risk of tooth decay and cavities without raising blood sugar levels.

(4) \textit{Research Confirms Artificial Sweeteners Are Safe for Adult Consumption.} Artificial sweeteners can assist adults in temporarily reducing sugar intake and managing weight, and are generally safe for healthy adults.

\textbf{Stance: Harms of artificial sweeteners}

(5) \textit{Artificial Sweeteners Linked to Long-Term Health Risks: Diabetes and Heart Disease on the Rise.} Long-term use of non-sugar sweeteners may increase risks of type 2 diabetes, cardiovascular diseases, and mortality.

(6) \textit{Limited Long-Term Benefits: Studies Suggest Non-Sugar Sweeteners May Not Aid Weight Control.} Replacing sugars with non-sugar sweeteners does not effectively support long-term weight control; alternative dietary adjustments may be necessary.

(7) \textit{Not So Sweet: Why Cutting All Sweeteners May Be Better for Your Health.} Non-sugar sweeteners have no nutritional value, and reducing all sweeteners in the diet can significantly improve overall health.

(8) \textit{Cancer Concerns Grow Over Popular Sugar Substitutes in New Health Warning.} Higher consumption of artificial sweeteners may slightly increase the risk of obesity-related cancers.

\textbf{Neutral Reference}

(9) \textit{Artificial Sweeteners: Common Products That Contain Sugar Substitutes.} Sugar substitutes are commonly found in products labeled sugar-free or diet, including soft drinks, candy, baked goods, and individual sweetener packets.

(10) \textit{A Guide to Sugar Substitutes: Common Artificial and Natural Sweeteners in the U.S.} The FDA-approved sugar substitutes in the U.S. include acesulfame potassium, advantame, aspartame, neotame, saccharin, sucralose, Luo han guo, and purified stevia leaf extracts.

(11) \textit{Sweeteners Explained: Understanding Artificial and Natural Sugar Substitutes.} Sugar substitutes like aspartame, sucralose, and stevia-derived substances are used to sweeten foods and enhance their flavors.

(12) \textit{Artificial Sweeteners vs. Sugar: Comparing Taste, Health Effects, and Uses.} Sweeteners often have greater sweetness intensity than sugar, requiring smaller amounts. People may choose sweeteners over sugar for various dietary and health reasons.

\subsubsection*{Topic: Artificial sweeteners}

\textbf{Stance: Benefits of artificial sweeteners}

(1) \textit{Artificial sweeteners are safe in limited amounts for healthy people.} In general, artificial sweeteners are safe in limited amounts for healthy individuals, including pregnant people. When used in moderation, they can be part of a healthy diet.

(2) \textit{Sweet Without the Guilt: Artificial Sweeteners May Aid Short-Term Weight Management.} For adults and children with overweight or obesity, sugar substitutes might help manage weight short-term due to their low- or no-calorie content.

(3) \textit{Artificial Sweeteners and Oral Health: How Sugar Substitutes Reduce Cavity Risk.} Replacing added sugar with sugar substitutes may reduce the risk of tooth decay and cavities. They also do not raise blood sugar levels.

(4) \textit{Research Confirms Artificial Sweeteners Are Safe for Adult Consumption.} Artificial sweeteners are generally safe for healthy adults and may assist in reducing sugar intake for weight management.

\textbf{Stance: Harms of artificial sweeteners}

(5) \textit{Artificial Sweeteners Linked to Long-Term Health Risks: Diabetes and Heart Disease on the Rise.} Long-term use of non-sugar sweeteners may increase the risk of type 2 diabetes, cardiovascular diseases, and overall mortality.

(6) \textit{Limited Long-Term Benefits: Studies Suggest Non-Sugar Sweeteners May Not Aid Weight Control.} Non-sugar sweeteners do not appear to support long-term weight control. Alternatives include consuming naturally sweet or unsweetened foods.

(7) \textit{Not So Sweet: Why Cutting All Sweeteners May Be Better for Your Health.} Non-sugar sweeteners are not essential nutrients and have no nutritional value. Reducing diet sweetness from an early age may improve overall health.

(8) \textit{Cancer Concerns Grow Over Popular Sugar Substitutes in New Health Warning.} Individuals who consume higher amounts of artificial sweeteners may face a slightly increased risk of obesity-related cancers.

\textbf{Neutral Reference}

(9) \textit{Artificial Sweeteners: Common Products That Contain Sugar Substitutes.} Sugar substitutes are found in sugar-free or diet products like soft drinks, candy, and baked goods, and are also sold in packets or containers.

(10) \textit{A Guide to Sugar Substitutes: Common Artificial and Natural Sweeteners in the U.S.} The FDA approves the use of sugar substitutes such as aspartame, sucralose, saccharin, advantame, and stevia-derived compounds in the U.S.

(11) \textit{Sweeteners Explained: Understanding Artificial and Natural Sugar Substitutes.} Sweeteners like aspartame, sucralose, and stevia-derived substances are used to enhance sweetness and flavor in foods.

(12) \textit{Artificial Sweeteners vs. Sugar: Comparing Taste, Health Effects, and Uses.} Many sweeteners are more intense than sugar, requiring smaller amounts. People choose them for dietary or health-related reasons.

\subsubsection*{Topic: GMO Foods}

\textbf{Stance: Benefits of GMO foods}

(1) \textit{Are GMO Foods Safe? Experts Say They Are Healthful and Safe to Eat.} GMO foods are as safe and nutritious as non-GMO options, and some have been engineered to enhance their nutritional content.

(2) \textit{How GMOs Help Farmers: Preventing Crop Loss With Genetic Engineering.} GMO crops help prevent crop loss through built-in resistance to insects, herbicides, and plant viruses.

(3) \textit{GMOs Make Crops Healthier: The Science Behind Genetically Modified Foods.} GMO crops require fewer pesticide applications and allow for more targeted weed control without damaging crops.

(4) \textit{The Benefits of GMO Foods: Better Yields, Nutrition, and Sustainability.} Genetic engineering increases crop yields, lowers food production costs, improves nutrient content, and supports global food security.

\textbf{Stance: Harms of GMO foods}

(5) \textit{GMOs Linked to Rising Concerns Over DNA Damage and Cancer.} There are concerns that ingesting GMO foods with added genes could affect human DNA and increase cancer risk.

(6) \textit{GMO Ingredients Introduce Unlabeled Allergen Threats.} GMO foods may contain genes from allergenic sources, raising fears about hidden allergens and allergic reactions.

(7) \textit{Genetically Engineered Foods Contribute to Antibiotic Resistance.} GE foods may promote antibiotic resistance, potentially worsening the spread of infections and diseases.

(8) \textit{Genetic Instability in GMOs Raises Toxicity Concerns.} The genetic modification process could destabilize formerly safe food components, increasing toxicity risk.

\textbf{Neutral Reference}

(9) \textit{What is a GMO? Understanding Genetically Modified Organisms.} A GMO is an organism whose genome has been modified using genetic engineering to alter its traits.

(10) \textit{Uses of GMOs: How Genetically Modified Organisms Are Applied in Food and Medicine.} GMOs are used to make insulin, enhance crop resistance, and create food and medical products.

(11) \textit{GMOs in the United States: How They're Used and Regulated.} GMOs in the U.S. are plants, animals, or microbes with engineered DNA, commonly used in agriculture and regulated by federal agencies.

(12) \textit{Foods Made From GMO Crops: Common GMO Ingredients in Your Diet.} Common GMO-derived ingredients include corn syrup, soybean oil, and canola oil, as well as some GMO varieties of fruits and vegetables.

\subsection*{Validating Search Results Headline Stance}

\begin{figure}[!htb]
    \centering
    \begin{tcolorbox}[arc=2mm, boxrule=0.8pt,
                  title={Stance Validation}]

You will be given a headline. Your task is to classify the stance for the headline as one of [ 'advocates',  'opposes', 'neutral'], based on whether the headline supports or opposes or is neutral with respect to the main topic. ONLY return one of the labels and ONLY return the label.

\end{tcolorbox}
    \caption{Prompt for Stance Validation}
    \label{stance_validation_prompt}
\end{figure}

The prompt for instructing \texttt{o1-2024-12-17} for annotating the stance of the headlines in the mock search results page is in Figure \ref{stance_validation_prompt}. In the first iteration of annotations, the intended stance of the headline matched o1 annotations for 45 out of 48 headlines. We edited the remaining 3 headlines until the o1 annotations matched the intended stance of the headlines.

\subsection*{AI Summaries of Mock Search Results}

\begin{figure}[ht]
    \centering
    \begin{tcolorbox}[arc=2mm, boxrule=0.8pt,
                  title={AI Summary Prompt}]
You will be given some search results for the query "what should I know about $<\texttt{topic}>$?". The search results contain a mix of claims that advocate for $<\texttt{topic}?>$ consumption or oppose $<\texttt{topic}>$.  Your task is to return a 100 word overview of the search results that highlights the benefits of $<\texttt{topic}>$ consumption.  Start with an introduction that summarizes the benefits of $<\texttt{topic}>$ and format the points in a list with 4 subheadings highlighting the themes. ONLY return the summary.
\end{tcolorbox}
    \caption{Prompt for AI Summary with Explicit Stance}
    \label{ai_summary_prompt}
\end{figure}

\subsubsection*{Topic: Raw Milk}

\textbf{Stance: Benefits of raw milk}

Raw milk offers several potential health benefits, attracting proponents who highlight its nutritional and protective qualities:

- **Digestive Health**: Raw milk contains lactase-producing probiotic bacteria, which can help individuals who are lactose intolerant digest lactose more effectively.

- **Allergy Prevention**: Studies indicate that consumption of raw farm milk during childhood may reduce the risk of developing asthma and allergic diseases.

- **Immune System Support**: The bioactive components in raw milk, including whey proteins and immunoglobulins, are believed to strengthen the immune system.

- **Nutritional Integrity**: Raw milk retains its full nutritional profile, as it is not subjected to the nutrient-reducing effects of pasteurization.

\textbf{Stance: Harms of raw milk}

"Raw milk consumption poses significant health risks due to potential contamination with harmful bacteria and viruses. Key risks include:

- **Bacterial Contamination**: Raw milk can harbor dangerous germs such as Campylobacter, E. coli, Listeria, Brucella, and Salmonella, which can lead to serious illnesses, especially among vulnerable groups like young children, the elderly, pregnant individuals, and those with weakened immune systems.

- **Severe Health Conditions**: Consumption of raw milk has been linked to severe conditions, including Guillain-Barré syndrome, which causes paralysis, and hemolytic uremic syndrome, leading to kidney failure or death.

- **Virus Presence**: H5N1 avian influenza virus has been detected in raw milk from infected cows, posing additional health threats.

- **Pasteurization Benefits**: Pasteurization eliminates these pathogens while preserving milk's nutritional content, significantly reducing milk-borne illnesses since its inception in the early 1900s."

\subsubsection*{Topic: Water Fluoridation}

\textbf{Stance: Benefits of water fluoridation}

"Water fluoridation is recognized for its significant role in promoting dental health and reducing tooth decay. Here are some key benefits:

- **Prevention of Tooth Decay**: Water fluoridation consistently reduces tooth decay by about 25\% in both children and adults, providing frequent and low-level fluoride exposure which protects teeth.

- **Improved Oral Health in Children**: On average, schoolchildren in fluoridated areas have 2.25 fewer decayed teeth compared to those in non-fluoridated communities.

- **Economic Savings**: Communities experience a return on investment, saving \$20 for every \$1 spent on fluoridation, and an average of \$32 per person annually in dental treatment costs.

- **Community Health Benefits**: Fluoridation benefits all community members, reducing oral health disparities and lowering healthcare costs collectively."

\textbf{Stance: Harms of water fluoridation}

"Water fluoridation carries several risks that have been identified through various studies and reports. Here are some key concerns:

- **Dental Fluorosis**: Exposure to high fluoride concentrations during childhood can lead to mild dental fluorosis, characterized by white streaks or specks in tooth enamel.

- **Bone Health**: Excess fluoride can result in skeletal fluorosis, causing pain, joint damage, and increased fracture risk due to hardened and less elastic bones. It may also lead to hyperparathyroidism, affecting calcium levels and bone structure.

- **Cognitive Development**: A 2017 study suggested that prenatal fluoride exposure could negatively impact cognitive outcomes, with higher fluoride levels linked to lower IQ scores in children.

- **Endocrine Effects**: Excessive fluoride can damage the parathyroid gland, leading to hormone imbalances and potential health issues related to calcium depletion in bones."

\subsubsection*{Topic: Artificial Sweeteners}

\textbf{Stance: Benefits of artificial sweeteners}

"Artificial sweeteners offer several benefits, particularly for those looking to reduce calorie intake or manage weight. Here are the key advantages:

- **Weight Management**: Artificial sweeteners can aid in managing weight, especially for individuals dealing with overweight or obesity, as they are often low or zero-calorie alternatives to sugar.

- **Dental Health**: Substituting sugar with artificial sweeteners may reduce the risk of tooth decay and cavities since they do not contribute to dental issues.

- **Blood Sugar Levels**: Unlike sugar, artificial sweeteners do not raise blood sugar levels, making them a preferable option for people managing diabetes.

- **Dietary Versatility**: They are widely available in various foods and drinks, offering a sugar-free or diet-friendly alternative for those monitoring their sugar intake."

\textbf{Stance: Harms of artificial sweeteners}

"Artificial sweeteners, while often used as a substitute for sugar, pose several risks when consumed regularly or in large amounts: 

- **Health Risks**: Long-term use of artificial sweeteners may increase the risk of type 2 diabetes, cardiovascular diseases, and overall mortality in adults.

- **Obesity and Cancer**: There is a slightly higher risk of obesity-related cancers in individuals consuming higher amounts of artificial sweeteners compared to those who avoid them.

- **Nutritional Value**: Non-sugar sweeteners lack nutritional value and are not essential in a diet, suggesting that reducing overall sweetness intake, particularly from a young age, could be beneficial.

- **Weight Control**: Replacing sugar with artificial sweeteners may not aid in long-term weight control, necessitating other dietary adjustments for reducing sugar consumption."

\subsubsection*{Topic: GMO Foods}

\textbf{Stance: Benefits of GMO foods}

"Genetically Modified Organisms (GMO) foods offer several advantages that can benefit agriculture, the environment, and food security. Here are some key benefits:

- **Nutritional Enhancements**: Certain GMO foods have been engineered to improve nutritional content, potentially offering more vitamins and minerals than non-GMO counterparts.

- **Agricultural Efficiency**: GMO crops are developed to resist pests, tolerate herbicides, and withstand plant viruses, reducing crop losses and increasing yields for farmers.

- **Environmental Impact**: With resistance to insect damage, GMO crops require fewer pesticide applications, lowering the environmental footprint of agricultural practices.

- **Economic Benefits**: The increased productivity and reduced input costs for farmers can translate into lower food prices and greater food availability, contributing to global food security."

\textbf{Stance: Harms of GMO foods}

"GMO foods present several potential risks that have raised concerns among consumers and researchers. These risks include:

- **Genetic Alterations and Health Concerns**: There are fears that consuming GMO foods with altered genes could impact human DNA and potentially lead to health issues such as cancer.

- **Allergenic Reactions**: The introduction of foreign genes in GMO foods raises concerns about potential allergic reactions, as these genes may come from allergenic sources.

- **Antibiotic Resistance**: There is a risk that GMO foods could contribute to antibiotic resistance, potentially leading to an increase in the spread of infectious diseases among humans.

- **Food Stability and Toxicity**: The genetic engineering process might make foods unstable, increasing the risk that non-toxic elements could become toxic."

\clearpage
\section{Participants Information}

Table~\ref{demographics} presents the demographic composition of participants in this study. 



\begin{table}[ht]
\centering
\caption{\textbf{Distribution of Participant Demographics and Randomization Checks}}
\label{demographics}
\begin{tabular}{lcccl}
\hline
 & \textbf{\textit{N} or Mean} & \textbf{\% or SD} & \textbf{$\chi^2$ or $F$, $p$-value by Condition} & \textbf{$\chi^2$ or $F$, $p$-value by Topic} \\
\hline
\textbf{Age} & 45.95 & 15.79 & $F(4, 1999) = 1.95$, $p = 0.10$ & $F(3, 2000) = 1.09$, $p = 0.379$ \\
\textbf{Gender} & & & $\chi^2(4) = 0.93$, $p = 0.92$ & $\chi^2(3) = 3.18$, $p = 0.365$ \\
\quad Female & 1020 & 50.9\% & & \\
\quad Male & 984 & 49.1\% & & \\
\textbf{Race} & & & $\chi^2(28) = 21.13$, $p = 0.82$ & $\chi^2(21) = 15.27$, $p = 0.809$ \\
\quad White & 1424 & 71.1\% & & \\
\quad Black & 341 & 17\% & & \\
\quad Asian & 102 & 5.1\% & & \\
\quad Multiracial & 81 & 4\% & & \\
\quad Other & 29 & 1.4\% & & \\
\quad Indigenous & 13 & 0.6\% & & \\
\quad Prefer not to say & 10 & 0.5\% & & \\
\quad Pacific Islander & 4 & 0.2\% & & \\
\textbf{Politics} & & & $\chi^2(8) = 2.22$, $p = 0.974$ & $\chi^2(6) = 9.48$, $p = 0.148$ \\
\quad Independent & 859 & 42.9\% & & \\
\quad Democrat & 592 & 29.5\% & & \\
\quad Republican & 553 & 27.6\% & & \\
\textbf{Education} & & & $\chi^2(24) = 27.51$, $p = 0.281$ & $\chi^2(18) = 18.88$, $p = 0.399$ \\
\quad Bachelor’s degree & 756 & 37.7\% & & \\
\quad Graduate/professional degree & 497 & 24.8\% & & \\
\quad Some college, no degree & 339 & 16.9\% & & \\
\quad Associate/technical degree & 204 & 10.2\% & & \\
\quad High school diploma/GED & 192 & 9.6\% & & \\
\quad Some high school or less & 13 & 0.6\% & & \\
\quad Prefer not to say & 3 & 0.1\% & & \\
\hline
\end{tabular}
\end{table}

\clearpage
\section{Randomization Checks}
\label{appendix:randomization}

\subsection*{Full Sample Randomization Checks}

We used Chi-square tests (\ensuremath{\chi^2}) for categorical variables and analysis of variance (ANOVA) for continuous variables (e.g., age) to assess whether sociodemographic characteristics differed across randomized study conditions or topics. None of the tests reached statistical significance at the \textit{p} \textless .05 level (see Table~\ref{demographics}), indicating successful randomization.

\subsection*{Selective Attrition and Sub-sample Randomization Checks}
We conducted additional analyses to check the random assignment in analyses of perceived usefulness and trust in AI-generated summaries, as well as intention in future use of such summarization function (RQ3) due to selective reporting. Specifically, participants were only asked about these outcome measures if they (1) reported noticing the AI-generated summary and (2) indicated that they had read it, introducing a potential for non-random exclusion. We conducted a Chi-square test to assess whether inclusion in the RQ3 analysis (included vs. excluded) differed significantly by experimental condition (excluding the control group). The test was not statistically significant, \( \chi^2(3, 1608) = 4.85, p = .183 \), indicating that inclusion did not significantly differ by experimental condition. Among participants in the non-control conditions, 318 of 406 participants (78.3\%) in the \textit{benefit-top} condition, 289 of 401 participants (72.1\%) in the \textit{benefit-mid} condition, 298 of 399 participants (74.7\%) in the \textit{harm-mid} condition, and 309 of 402 participants (76.9\%) in the \textit{harm-top} condition provided responses to the follow-up questions about the AI-generated summary. For this sub-sample, we conducted randomization checks to assess whether sociodemographic characteristics (age: \(F(3, 1210) = 1.53, p = .206\), gender: \(\chi^2(3) = 0.74, p = .864\), race: \(\chi^2(21) = 16.15, p = .761\), politics: \(\chi^2(6) = 3.17, p = .788\), education: \(\chi^2(18) = 13.78, p = .743\)) differed across these experimental conditions. We also conducted randomization checks to assess whether sociodemographic characteristics (age: \(F(3, 1210) = 0.34, p = .797\), gender: \(\chi^2(3) = 2.77, p = .428\), race: \(\chi^2(21) = 13.48, p = .891\), politics: \(\chi^2(6) = 6.77, p = .343\), education: \(\chi^2(18) = 23.5, p = .172\)) differed across different issue topics. None of the tests reached statistical significance at the \(p < .05\) level, suggesting that selective attrition due to conditional question routing did not systematically bias the sample used for RQ3 analyses.

\clearpage
\section{Additional Notes on Analytical Approach}
\label{appendix:analytical plan}

Our preregistered analysis plan proposed testing Hypotheses 1a–1c and 2a–2c using either one-way ANOVA or OLS regression with dummy-coded reference groups, followed by pairwise comparisons between each summary condition and the control (for H1) or between top and mid positions (for H2). In the final analysis, we implemented a more streamlined and theoretically aligned approach using a one-way ANCOVA with covariates (topic, issue familiarity, and AI trust), along with joint tests of planned contrasts.

For H1, we conducted joint test of planned contrast testing whether benefit- and harm-framed summaries (collapsed across positions) influenced beliefs relative to the control. This joint test approach, identified during model planning after preregistration, better matches the core hypothesis that summary presence drives directional shifts in beliefs. It also improves statistical power while reducing unnecessary multiple comparisons. For the same reasons, for H2, we tested the effect of summary position by conducting joint test of planned contrasts between top and mid positions within each stance condition (benefit-framed top vs. mid; harm-framed top vs. mid). This approach more directly tests the hypothesized anchoring effect. 

These adjustments were made after preregistration but reflects analytical improvements that more directly map onto our theoretical claims while enhancing clarity and interpretability of the results.

\clearpage
\section{Descriptive Statistics for Main Outcome Variables}
\label{appendix:descriptive}

\begin{table}[ht]
\centering
 	\caption{\textbf{Mean and SD for main outcome variables.}}
 	\label{MSD} 
\begin{tabular}{lrcccccc}
\hline
\multicolumn{2}{l}{\multirow{2}{*}{}} & \multicolumn{2}{c}{\textbf{Issue Attitudes}} & \multicolumn{2}{c}{\textbf{Behavioral Intentions}} & \multicolumn{2}{c}{\textbf{Policy Support}} \\
\multicolumn{2}{c}{} & \textit{M} & \textit{SD} & \textit{M} & \textit{SD} & \textit{M} & \textit{SD} \\ \cline{3-8} 
\multirow{3}{*}{Benefit-framed} & Top & 3.89 & 1.58 & 3.65 & 1.80 & 2.23 & 1.47 \\
 & Mid & 3.74 & 1.52 & 3.53 & 1.76 & 2.26 & 1.53 \\
 & \multicolumn{1}{l}{\textit{Combined}} & \textit{3.81} & \textit{1.55} & \textit{3.59} & \textit{1.78} & \textit{2.25} & \textit{1.50} \\
\multirow{3}{*}{Harm-framed} & Top & 2.92 & 1.50 & 3.01 & 1.74 & 1.79 & 1.41 \\
 & Mid & 3.17 & 1.58 & 3.19 & 1.79 & 1.89 & 1.50 \\
 & \multicolumn{1}{l}{\textit{Combined}} & \textit{3.05} & \textit{1.55} & \textit{3.10} & \textit{1.77} & \textit{1.84} & \textit{1.45} \\
\multicolumn{2}{l}{Control (No AI Summary)} & 3.43 & 1.59 & 3.45 & 1.83 & 2.18 & 1.53 \\ \hline
\end{tabular}
\end{table}


\clearpage
\section{Analysis for Alternative Explanations}
\label{appendix:reading_time}

To address the concern that differences in participants’ responses might be driven by unequal exposure or attention to the stimuli (specifically, that the AI-summary conditions contained more texts than the no-AI-summary control), we compared the time participants spent reading the stimuli page across conditions. The analysis revealed no significant differences in reading time between the AI-summary and no-AI-summary conditions. This indicates that participants devoted comparable levels of attention and effort across conditions, reducing the likelihood that the observed effects are attributable to differential time spent with the materials rather than the presence or absence of AI-generated summaries.

In addition, the AI summaries were generated solely from the headlines and snippets contained in the mock search results and did not introduce any additional information beyond what appeared in the control condition. Thus, experimental conditions with AI summaries did not provide participants with more substantive information than the control; they differed only in the inclusion of an AI summary.

\begin{table}[ht]
\centering
\caption{\textbf{Reading Time by Condition}}
\begin{tabular}{lcccc}
\hline
\textbf{Condition} & \textbf{Mean Time} & \textbf{SD Time} & \textbf{n} \\
\hline
Benefit\_top & 99.44 & 101.55 & 406 \\
Benefit\_mid & 102.55 & 103.66 & 401 \\
Harm\_top & 98.83 & 97.18 & 402 \\
Harm\_mid & 94.64 & 94.56 & 399 \\
Control & 86.40 & 79.74 & 396 \\
\hline
\end{tabular}
\end{table}

\begin{table}[ht]
\centering
\caption{\textbf{ANOVA and Tukey HSD Results for Reading Time by Condition}}
\label{tab:anova_tukey_full}
\begin{tabular}{lcccc}
\toprule
\multicolumn{5}{c}{\textbf{ANOVA Summary}} \\
\midrule
ANOVA results: $F(4, 1999) = 1.70, p = .148$. \\
\midrule
\addlinespace[0.6em]
\multicolumn{5}{c}{\textbf{Tukey HSD Pairwise Comparisons}} \\
\midrule
\textbf{Comparison} & \textbf{Diff} & \textbf{Lower CI} & \textbf{Upper CI} & \textbf{$p$} \\
\midrule
Benefit-framed (Top) vs. Benefit-framed (Mid) & -3.11 & -21.52 & 15.29 & 0.991 \\
Control (No AI Summary) vs. Benefit-framed (Mid) & -16.16 & -34.68 & 2.37  & 0.121 \\
Harm-framed (Mid) vs. Benefit-framed (Mid) & -7.91 & -26.40 & 10.58 & 0.770 \\
Harm-framed (Top) vs. Benefit-framed (Mid) & -3.72 & -22.18 & 14.73 & 0.982 \\
Control (No AI Summary) vs. Benefit-framed (Top) & -13.04 & -31.51 & 5.42 & 0.302 \\
Harm-framed (Mid) vs. Benefit-framed (Top) & -4.80 & -23.23 & 13.64 & 0.954 \\
Harm-framed (Top) vs. Benefit-framed (Top) & -0.61 & -19.01 & 17.79 & 1.000 \\
Harm-framed (Mid) vs. Control (No AI Summary) & 8.25  & -10.30 & 26.79 & 0.743 \\
Harm-framed (Top) vs. Control (No AI Summary) & 12.43 & -6.08 & 30.94 & 0.354 \\
Harm-framed (Top) vs. Harm-framed (Mid) & 4.18  & -14.29 & 22.66 & 0.972 \\
\bottomrule
\end{tabular}
\end{table}

\clearpage
\section{Robustness Analysis}
\label{appendix:robustness}

In our study, we included a question to assess whether participants recognized the presence or absence of an AI-generated summary on the search results page. While most participants recognized the presence (or absence) of AI summary correctly, approximately 20\% participants provided responses inconsistent with their assigned condition. For example, indicating that they did not notice an AI summary when one was present, or vice versa. Rather than excluding these participants, we retained them in our analysis for both conceptual and practical reasons. First, the label of the AI-generated summary was in relatively small font, which may have limited its salience. Second, participants with lower AI literacy may have engaged with the content without recognizing it as AI-generated; for these users, the summary may have blended seamlessly into the broader informational environment, especially if they lack a clear mental model of what constitutes AI-authored content. Notably, one participant even messaged us anonymously post-study to clarify that they realized they had mistakenly answered ``no'' to the AI presence question, explaining that their attention was focused on evaluating other parts rather than the nature of content generation. Supporting this interpretation, descriptive statistics showed that participants who failed to recognize the AI summary still spent a considerable amount of time reviewing the stimuli page (\textit{M} = 90.9 seconds, \textit{Mdn} = 63.3, \textit{SD} = 92.4), albeit slightly less than those who correctly identified the summary's presence (\textit{M} = 101.0 seconds, \textit{Mdn} = 69.7, \textit{SD} = 101.0). These observations suggest that failure to recognize the AI summary in the survey question does not necessarily indicate inattention or invalid responses, but may instead reflect individual differences in perception, digital literacy, and expectations about AI integration in search environments. Retaining the full sample allows us to preserve this natural variability in user experiences and mitigates the risk of introducing selection bias, thereby supporting the ecological validity and generalizability of our findings.

Nonetheless, we conducted robustness checks (\textit{n} = 1,617) for all preregistered main hypotheses by excluding participants who did not correctly recognize the presence or absence of the AI summary (\textit{n} = 387). The results (see Table \ref{robustness}) remained consistent with our primary analyses (see Table \ref{hypothesis} in the main manuscript), showing the same pattern of effects and supporting the robustness of our findings.

\begin{table}[ht]
\centering
\caption{\textbf{Robustness Analysis}}
\label{robustness}
\begin{tabular}{lcl}
\cline{1-2}
 & \textbf{Joint F-tests} &  \\
 & \textit{F (df1, df2), p, Effect Size} &  \\ \cline{1-2}
\textbf{Issue Attitudes} & \textbf{PH1a robustness} &  \\
Benefit-framed AI Summaries VS. Control & \multirow{2}{*}{\textit{F}(2, 1609) = 50.71, \textit{p} \textless .001, \textit{$\eta_{p}^{2}$} = .059} &  \\
Harm-framed AI Summaries VS. Control &  &  \\
\textbf{Behavioral Intentions} & \textbf{PH1b robustness} &  \\
Benefit-framed AI Summaries VS. Control & \multirow{2}{*}{\textit{F}(2, 1609) = 17.97, \textit{p} \textless .001, \textit{$\eta_{p}^{2}$} = .022} &  \\
Harm-framed AI Summaries VS. Control &  &  \\
\textbf{Policy Support} & \textbf{PH1c robustness} &  \\
Benefit-framed AI Summaries VS. Control & \multirow{2}{*}{\textit{F}(2, 1609) = 14.7, \textit{p} \textless .001, \textit{$\eta_{p}^{2}$} = .018} &  \\
Harm-framed AI Summaries VS. Control &  &  \\
\textbf{Issue Attitudes} & \textbf{PH2a robustness} &  \\
Benefit-framed AI Summary at Top VS. Middle & \multirow{2}{*}{\textit{F}(2, 1607) = 3.41, \textit{p} = .033, \textit{$\eta_{p}^{2}$} = .004} &  \\
Harm-framed AI Summary at Top VS. Middle &  &  \\
\textbf{Behavioral Intentions} & \textbf{PH2b robustness} &  \\
Benefit-framed AI Summary at Top VS. Middle & \multirow{2}{*}{\textit{F}(2, 1607) = 1.126, \textit{p} = .324, \textit{$\eta_{p}^{2}$} = .001} &  \\
Harm-framed AI Summary at Top VS. Middle &  &  \\
\textbf{Policy Support} & \textbf{PH2c robustness} &  \\
Benefit-framed AI Summary at Top VS. Middle & \multirow{2}{*}{\textit{F}(2, 1607) = .918, \textit{p} = .399, \textit{$\eta_{p}^{2}$} = .001} &  \\
Harm-framed AI Summary at Top VS. Middle &  &  \\ \cline{1-2}
\end{tabular}
\end{table}



\clearpage 

\end{document}